\documentclass[doublecolumn]{IEEEtran}
\usepackage{subfigure}
\usepackage{booktabs}
\usepackage{bm}
\usepackage{multirow}
\usepackage{array}
\usepackage{amsfonts}
\usepackage{float}
\usepackage{color}
\usepackage[ruled,linesnumbered]{algorithm2e}

\usepackage{cite}

\ifCLASSINFOpdf
 \usepackage[pdftex]{graphicx}

\usepackage{amsmath}

\usepackage{algorithmic}

\usepackage{array}

\begin{document}

%
\title{A Convolutional Neural Network with Mapping Layers for Hyperspectral Image Classification}
%
%
%

\author{Rui~Li,
        Zhibin~Pan, Yang~Wang, and Ping~Wang
        \thanks{This work is supported in part by Zhejiang Provincial Natural Science Foundation of China under Grant No. LQY19F010001.}
\thanks{Rui~Li is with the School of Electronic and Information Engineering, Xi'an Jiaotong University, Xi'an, P. R. China, 
710049.}
\thanks{Zhibin~Pan is with the School of Electronic and Information Engineering, Xi'an Jiaotong University, Xi'an, P. R. China, 710049, and also with the Research Institute of Xi'an Jiaotong University, Zhejiang, P. R. China, 311215,
 (e-mail: zbpan@mail.xjtu.edu.cn).}
\thanks{Yang~Wang is with the School of Electronic and Information Engineering, Xi'an Jiaotong University, Xi'an, P. R. China, 710049.}
\thanks{Ping~Wang is with the School of Electronic and Information Engineering, Xi'an Jiaotong University, Xi'an, P. R. China, 710049, and also with the Research Institute of Xi'an Jiaotong University, Zhejiang, P. R. China, 311215.}
}

%
%

\markboth{Under review on IEEE Trans. Geosci. Remote Sen.}%
{Shell \MakeLowercase{\textit{et al.}}: Bare Demo of IEEEtran.cls for IEEE Journals}
%



\maketitle

\begin{abstract}
In this paper, we propose a convolutional neural network with mapping layers (MCNN) for hyperspectral image (HSI) classification. The proposed mapping layers map the input patch into a low dimensional subspace by multilinear algebra. We use our mapping layers to reduce the spectral and spatial redundancy and maintain most energy of the input. The feature extracted by our mapping layers can also reduce the number of following convolutional layers for feature extraction. Our MCNN architecture avoids the declining accuracy with increasing layers phenomenon of deep learning models for HSI classification and also saves the training time for its effective mapping layers. Furthermore, we impose the 3-D convolutional kernel on convolutional layer to extract the spectral-spatial features for HSI.
We tested our MCNN on three datasets of Indian Pines, University of Pavia and Salinas, and we achieved the classification accuracy of 98.3\%, 99.5\% and 99.3\%, respectively. Experimental results demonstrate that the proposed MCNN can significantly improve the classification accuracy and save much time consumption.
\end{abstract}


%
\IEEEpeerreviewmaketitle

\section{Introduction}
\IEEEPARstart{H}{yperspectral} image (HSI) has drawn much attention in recent years. HSI processing is widely used in various remote sensing fields, such as land use analysis, urban planning and environment monitoring\cite{khan2018modern,fan2018semi,kang2015extended}. In order to alleviate high spectral dimension in HSI classification, many dimension reduction methods are proposed. Generally, they can be divided into two categories: feature selection methods and feature extraction methods. The feature selection methods aim at preserving the most representative bands and discarding those making no contributions to the classification results. Wang \cite{wang2016salient} adopts manifold ranking as an unsupervised feature selection method to choose the most representative bands.
Du and Yang propose a similarity-based unsupervised band selection method\cite{du2008similarity}.
 Moreover, the multitask joint sparse representation-based method\cite{yuan2016hyperspectral} integrates band selection method with Markov random field. 
However, feature selection methods can only select existing bands from HSI. Thus, the feature extraction methods are designed to find a best method to map the original high-dimensional feature into a low-dimensional subspace while keeping the most useful dimensions for classification. The most representative methods are principal component analysis (PCA)\cite{jiang2018superpca} and its variants\cite{prasad2008limitations,hossain2011unsupervised,laparra2015dimensionality}. 
The linear discriminant analysis is also a widely used feature extraction method which searches the optimal projection subspace by maximizing between-class scatter matrix and minimizing within-class scatter matrix \cite{bandos2009classification}. Apart from this, some graph-based feature extraction methods are proposed for its similarity to model relations among the objects \cite{zhai2016modified}\cite{wong2012discover}. 
In this paper we focus on the feature extraction method in order to improve the HSI classification performance.

The complex spatial and spectral distributions make it difficult to classify the objects in HSI, a good methodology to incorporate the spatial feature and spectral feature for HSI classification is the most concerned issue in this field. In Wang's locality and structure regularized low rank representation (LSLRR) model \cite{wang2018locality}, a new distance metric is introduced to combine both spatial and spectral features. However, the spectral bands in HSI are much more redundant than its spatial information, thus many traditional methods extract the most discriminative spatial features or bands and use them to train the classifiers. Support vector machine (SVM) is the most commonly used classifier in these traditional HSI classification methods. SVM-based classifiers are state-of-the-art methods with excellent classification accuracy for a long time. Recently, SVM is combined with edge-preserving filter to derive the spectral-spatial information for classification\cite{kang2014spectral}. Not only suffering from curse of dimensionality, HSI classification also subjects to the problem of large spatial variability of spectral signature\cite{camps2005kernel}. To deal with the spatial variability of spectral signature, the spectral information is extracted for classification in many methods, such as extended morphological profiles\cite{benediktsson2003classification}, spectral-spatial kernels\cite{camps2006composite} and super-pixels\cite{galasso2012video}. In this paper, we also try to extract spectral-spatial feature for HSI classification.

For many HSI classification methods, they have a drawback that one or two layers of nonlinear transformation is not enough to represent the spatial and spectral features. The linear SVM and logistic regression can be viewed as single layer classifier, whereas decision tree and SVM with kernels are considered to have two layers\cite{bengio2013representation}. It is known that human brains perform well in tasks like object recognition for its multiple stages of processing from retina to cortex\cite{kruger2013deep}. In order to extract the better invariant feature of data, a classifier of multi-layer structure is advised. Therefore, deep architectures have been shown to yield promising performance in image classification task. Recently, with the development of deep learning, Chen first introduces deep learning to HSI classification\cite{chen2014deep}. After that, many other deep learning methods are applied in HSI classification to improve the performance \cite{kong2018spectral}\cite{feng2019cnn}\cite{xu2017multisource}. Wang employs recurrent convolutional network for scene classification of remote sensing\cite{wang2018scene}.
Among them, the convolutional neural network (CNN) has achieved promising performance in many HSI processing fields and many researchers have proved that CNN can deliver state-of-the-art performance in HSI classification\cite{chen2016deep}. Deep learning techniques are able to automatically learn hierarchical features (from low-level to high-level) from raw input data, and the reason is that scene classification can explore the salient features in remote sensing scenes. Such learned features have achieved tremendous success in many machine vision tasks. Although deep learning methods like stacked autoencoder (SAE)\cite{vincent2010stacked} and deep brief network (DBN)\cite{lee2009convolutional} can extract the spectral-spatial feature hierarchically in a layer-wise training method, the input image patch should be flattened as a vector to meet the input requirement that may break the inherent spectral-spatial structure in HSI. We recommend that the input patch stays as the data cube to remain as much original information as possible. Lee and Kwon\cite{lee2017going} incorporated residual learning with convolutional layers to form a contextual CNN, but spectral features and spatial features are processed in the same way which is intuitively incorrect. Meanwhile, the classification accuracies of the CNN models decrease when the networks become deeper. The spectral-spatial residual network (SSRN)\cite{zhong2018} is developed to overcome this drawback. SSRN contains shortcut connections between every other convolutional layers, thus, a robust representation can be learned from the original HSI for classification. 
According to the previous research, we try to design a new architecture to extract better spectral-spatial feature and improve the HSI classification performance.

Although many deep learning-based methods have been introduced to enhance the classification performance, HSI classification still suffers from the following three problems: (1) The spectral-spatial information in HSI should be fully utilized in CNN to improve the classification performance. A proper way to organize the spatial and spectral information is needed. (2) The classification accuracy of the CNN models decreases when the network becomes deeper. A more effective network with less layers is needed. (3) The training procedure of CNN models is time consuming, and the training samples usually go hundreds of training sessions to converge. All these problems limit the promotion of deep learning in HSI classification. 

To counter these drawbacks, we propose a convolutional neural network with mapping layers (MCNN) in this paper. As other deep networks extract the spectral-spatial features through multiple-layer mappings, the weights in all these layers are randomly initialized and updated by backward propagation. This leads to the problem that layers of network should go deep to extract the complicated and abstract features. Thus, an effective method to extract features of HSI may work better than the convolutional layers, especially when there is a large amount of spectral and spatial redundancies existing in HSI. By properly streamlining the redundancy and keeping the most significant part of energy in HSI, we can obtain effective HSI features. Despite the fact that deep convolutional layers can extract spectral-spatial features effectively, they also bring the problems of large computational cost and the classification accuracy decreasing with network going deeper. In this paper, we build a new kind of layer called mapping layer to effectively extract the feature and reduce the redundancy. 

The mapping layer can reduce the redundancy in specific mode of HSI tensor and keep the feature of the most part of the energy. The input of this kind of layer is the patch of HSI, and the data still retain the cube form. The spectral and spatial structures are kept, which are also beneficial to feature extraction. The weights of mapping layer are not updated by backward propagation, instead they are obtained by decomposition of input HSI patches. This weight setting method can obtain more suitable weights for input HSI than randomly initialized weights in convolutional layers. The weights in mapping layer are non-trainable, thus the mapping layer can save much more computational cost than convolutional layer because the weights in convolutional layers are updated by backward propagation in each training sessions. We adopt mapping layers to obtain better spectral-spatial feature and reduce the number of convolutional layers. Meanwhile the problem that classification accuracy decreases with the network going deeper is avoided as our network architecture only has a few layers. Considering the difference between HSI and natural image is that HSI has much more bands, the bands should not be the redundancy but be the key to improve HSI classification performance. To extract the finer spectral-spatial feature, a 3-D convolutional kernel is also adopted. The 3-D convolutional kernel can extract the differences between bands as features, and more spectral information can help us improve the classification performance of our network.

The highlights of this paper are listed below: 
\begin{itemize}
  \item [1)]
We build mapping layers to extract the spectral-spatial feature in HSI. We obtain the mapping kernels by decomposition of the input HSI patches, and the combination of these specialized layers for HSI with convolutional layers can extract better features for classification.
  \item [2)]
We propose a new network architecture called convolutional neural network with mapping layers (MCNN). Our MCNN contains three sections: mapping section, convolutional section and fully connected section. The architecture takes into account feature extraction and dimension reduction, deploying only a few layers in MCNN architecture can save the computational time and avoid the problem that accuracy decreases as network going deeper.
  \item [3)]
Finer 3-D feature extraction method is adopted in convolutional layers. To extract the spectral-spatial feature, the input HSI patch is convoluted with 3-D kernels. As a result, we can obtain the spatial and spectral features at the same time. 
\end{itemize}

The rest of this paper is organized as follows: Section II describes the related knowledge of tensor decomposition. We give our MCNN architecture in Section III. Then, the network configuration and experimental results are reported in Section IV.  Discussions about the proposed mapping layer and performance are also offered in Section IV. Finally, conclusions are drawn in Section V.
\\
\section{Tensor Preliminaries}
Tensor decomposition is a method which decomposes a tensor into one core tensor and factor matrices (projection matrices). The core tensor implies the connections among the vectors of factor matrices in different modes\cite{zhang2018high}. Two kinds of tensor decompositions are widely used: the canonical polyadic (CP) decomposition, also known as parallel factor (PARAFAC)\cite{harshman1970foundations} decomposition, and the Tucker decomposition (TD)\cite{de2000multilinear} also known as high-order singular value decomposition (HOSVD)\cite{tucker1966some}.
HSI is a data cube constructed by plenty of bands, therefore it is naturally a third-order tensor. We can apply Tucker decomposition to decompose the tensor into three factor matrices and a core tensor where most of the energy is concentrated, then we can use this property to design our mapping layer and obtain better spectral-spatial feature.

\begin{table}[H]
   \centering
   \caption{Notations About Tensor}
   \label{tab1}
   \begin{tabular}{p{0.3\columnwidth}<{\centering}p{0.55\columnwidth}<{\centering}}
   \toprule
\multicolumn{2}{c}{Notations used in this paper}\\  
    \midrule
    $\otimes$  & Kronecker product \\
    $\bm{\mathcal{X}}\in\mathbb{R}^{{I_1} \times {I_2} \times \dots\times {I_N}}$ & A\ $N_{th}$-order tensor \\
    $x( {{i_1},{i_2},...,{i_N}} )$& The entry of tensor   $\bm{\mathcal{X}}\in\mathbb{R}^{{I_1} \times {I_2} \times \dots\times {I_N}}$ \\
    $\textbf{X}_{(1)}$ & The mode-1 matricization of $\bm{\mathcal{X}}$\\
    $\textbf{X}^{I\times{JK}}$&A matrix of size $I\times{JK}$\\
    $\textbf{U}^{(i)}$ & $i_{th}$-order\\
    $\textbf{X}_{(i)}^k$ & The $k_{th}$ slice along mode-$i$ in tensor $\bm{\mathcal{X}}$\\
    $\textbf{X}_{::i}$ &The $i_{th}$ slice in tensor $\bm{\mathcal{X}}$\\
    \bottomrule
   \end{tabular}
\end{table}
We list the notations used in this paper in Table I. The order of the tensor is also known as mode. The first-order tensor is actually a vector and the second-order tensor is known as matrix. The mode-n matricization means flattening the tensor into a matrix by fixing the $N_{th}$-order index and varying other indices. For a third-order tensor $\bm{\mathcal{X}}\in\mathbb{R}^{{I} \times {J} \times {K}}$, there are three mode-n matricizations.
\begin{equation}
\textbf{X}_{(1)}=\textbf{X}^{(I\times JK)}=[\textbf{X}_{::1},\dots,\textbf{X}_{::K}],
\end{equation}
\begin{equation}
\textbf{X}_{(2)}=\textbf{X}^{(J\times KI)}=[\textbf{X}_{:1:},\dots,\textbf{X}_{:J:}],
\end{equation}
\begin{equation}
\textbf{X}_{(3)}=\textbf{X}^{(K\times IJ)}=[\textbf{X}_{1::},\dots,\textbf{X}_{I::}].
\end{equation}

The mode-1 Kiers matricization of a third-order tensor $\bm{\mathcal{X}}$  results in $\textbf{X}_{(1)}$, and the flattening process is shown in Fig. 1. $\textbf{X}_{(2)}$ and $\textbf{X}_{(3)}$ are the mode-2 and mode-3 Kiers matricization results, respectively. The matricizations in this paper are all Kiers matricizations.
\begin{figure}[tpb]
\centering
\includegraphics[width=0.45\textwidth]{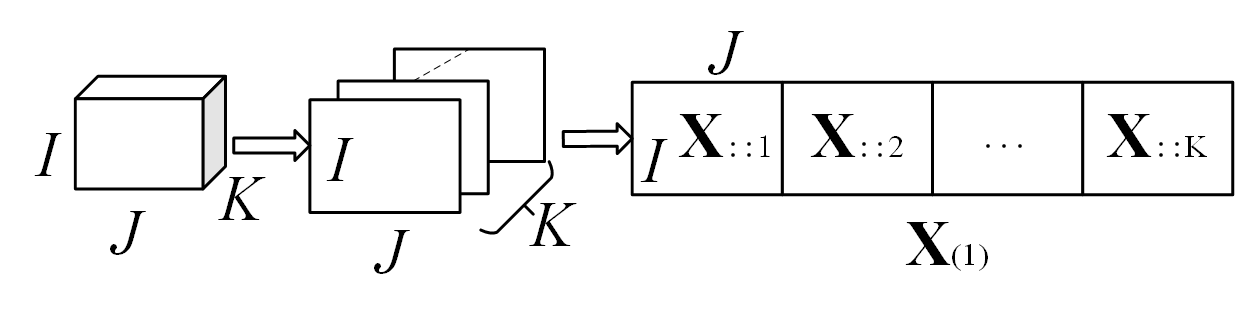}
\caption{The mode-1 matricization is obtained by flattening the tensor along its first mode.}
\label{fig1}
\end{figure}
The mode-n tensor-matrix product of a tensor $\bm{\mathcal{X}}\in\mathbb{R}^{{I_1} \times {I_2} \times \dots\times {I_N}}$ multiplied by matrix $\textbf{U}\in\mathbb{R}^{J_n\times{I_n}}$ is denoted as $\bm{\mathcal{X}}\times{_n\textbf{U}}$:
\begin{equation}
\bm{\mathcal{X}}\times_n\textbf{U}_{i_1,i_2,\dots,i_{n-1},i_n,i_{n+1},\dots,i_N}=\sum_{i_n=1}^{I_n}{x_{{i_1}{i_2}\dots{i_N}}\times u_{i_nj_n}},
\end{equation}
where the height of matrix $\textbf{U}$ should be the same as the mode-n size of tensor $\bm{\mathcal{X}}$. This multiplication changes the mode-n size of tensor $\bm{\mathcal{X}}$ from $I_n$ to $J_n$.

The $N_{th}$-order tensor can be resized to a new tensor with each mode resized as the height of the corresponding matrix. The equation is as below:
\begin{equation}
\bm{\mathcal{A}}=\bm{\mathcal{X}}\times{_1}\textbf{U}^{(1)}\times{_2}\textbf{U}^{(2)}\dots\times{_N}\textbf{U}^{(N)},
\end{equation}
where the width of matrix $\textbf{U}^{(n)} (n=1,2,\dots,N)$ should be equal to the size of mode-n of tensor $\bm{\mathcal{X}}$.

As the tensors in this paper are all third-order tensors, Tucker decomposition (TD) is introduced in version of third-order:
\begin{equation}
\bm{\mathcal{X}}=\bm{\mathcal{G}}\times{_1}\textbf{U}^{(1)}\times{_2}\textbf{U}^{(2)}\times{_3}\textbf{U}^{(3)},
\end{equation}
where $\bm{\mathcal{X}}$ is the original tensor, and $\bm{\mathcal{G}}$ is the core tensor of $\bm{\mathcal{X}}$, $\textbf{U}^{(1)},\textbf{U}^{(2)},\textbf{U}^{(3)}$ are factor matrices (projection matrices) where $\textbf{U}^{(n)}\textbf{U}^{(n)T}=I_{J_n}(n=1,2,3)$. The original tensor $\bm{\mathcal{X}}$ can be reconstructed by equation below:
\begin{equation}
\bm{\mathcal{G}}=\bm{\mathcal{X}}\times{_1}\textbf{U}^{(1)T}\times{_2}\textbf{U}^{(2)T}\times{_3}\textbf{U}^{(3)T}.
\end{equation}
\\
\section{The Proposed MCNN Architecture}
CNN is an efficient method for classification and feature extraction. The application of CNN on the task of HSI classification significantly improves the classification accuracy. However, it still suffers from the problems that the classification accuracy of the CNN model decreases when the network becomes deeper\cite{zhong2018}, and the training time is long for network with plenty of convolutional layers. In fact, all these drawbacks are caused by the increasing number of convolutional layers. The convolutional layers are adopted to extract spectral-spatial feature. Practically, two sequential $3\times3$ convolutional layers can perform better than a $5\times5$ convolutional layer \cite{simonyan2014very}. Therefore, many researchers adopt more and more convolutional layers in their networks, which lead to problems of decreasing classification accuracies and increasing computational time. Motivated by the analysis above,  we design a kind of more effective layer for HSI in this paper, aiming at reducing the convolutional layer and extracting better spectral-spatial feature. We will first introduce our mapping layer, and then give our architecture based on mapping layers and convolutional layers. In the end, we will talk about using 3-D convolutional kernels to extract spectral-spatial features.
\subsection{The Mapping Layer}
The input patch of HSI has spectral and spatial redundancy. In order to eliminate the redundancy and keep as much useful information for classification as possible, a special kind of layer for feature extraction is necessary. However, the convolutional layer only performs well on edge detection and texture recognition. Therefore, we newly design the mapping layer to extract the spectral-spatial feature and reduce the dimension.

For a third-order tensor, we need three mapping layers to extract energy concentrated features, each mapping layer maps a corresponding mode (order). To generate adaptive mapping layers for all input HSI patches, we apply our method on the averaged patch of all training samples. In this way, our mapping layer can reduce the common redundancy of all training patches and save the training time.
\begin{equation}
\mathop{\arg\min}_{\textbf{U}^{(1)},\textbf{U}^{(2)},\textbf{U}^{(3)}} \ \ \| \bm{\mathcal{X}}-f(\textbf{U}^{(1)},\textbf{U}^{(2)},\textbf{U}^{(3)})\|^2_2.
\end{equation}

Where $f(\textbf{U}^{(1)},\textbf{U}^{(2)},\textbf{U}^{(3)})=\bm{\mathcal{G}}\times{_1}\textbf{U}^{(1)}\times{_2}\textbf{U}^{(2)}\times{_3}\textbf{U}^{(3)}$.
Obtaining the three factor matrices is a coupling problem describled by (8), where $\bm{\mathcal{X}}\in\mathbb{R}^{{I_1} \times {I_2}\times {I_3}}$ is the input patch and $\textbf{U}^{(n)}\in\mathbb{R}^{{I_n} \times {R_n}}$ is the factor matrix. $R_n$ is the reduced dimensionality of factor matrix, which can reduce the redundant dimension of the mode according to (4).  A common strategy to optimize such a problem is the alternating least squares (ALS) method. We can solve only one factor matrix when other variables are fixed in an iteration. The optimal factor matrices are obtained after plenty of iterations.

We first initialize three factor matrices as $\{\textbf{U}^{(1)}_0,\textbf{U}^{(2)}_0,\textbf{U}^{(3)}_0\}$. The column vectors in initial factor matrices are orthogonal to each other. Then, the averaged HSI patch is flattened along three modes and three corresponding matricizations $\textbf{X}_{(1)},\textbf{X}_{(2)},\textbf{X}_{(3)}$ are obtained. We update the factor matrices according to (9)-(11) given below.

\begin{equation}
[\textbf{U}^{(1)},\textbf{V}_1,\textbf{T}_1]=SVD[\textbf{X}_{(1)}(\textbf{U}^{(3)}\otimes\textbf{U}^{(2)}),R_1],
\end{equation}
\begin{equation}
[\textbf{U}^{(2)},\textbf{V}_2,\textbf{T}_2]=SVD[\textbf{X}_{(2)}(\textbf{U}^{(1)}\otimes\textbf{U}^{(3)}),R_2],
\end{equation}
\begin{equation}
[\textbf{U}^{(3)},\textbf{V}_3,\textbf{T}_3]=SVD[\textbf{X}_{(3)}(\textbf{U}^{(2)}\otimes\textbf{U}^{(1)}),R_3].
\end{equation}

In each iteration, three factor matrices are updated alternatively. The final factor matrices are obtained when all of factor matrices can hardly be updated. These three factor matrices are used to construct three sequential mapping layers.

\begin{algorithm}[htb]
\caption{The training of mapping layers}
\KwIn{The training patches $\bm{\mathcal{X}}\in\mathbb{R}^{{I_1} \times {I_2} \times {I_3}}$ and reduced dimensionalities $R_1, R_2, R_3$.}
Initialized the factor matrix $\{\textbf{U}^{(1)}_0\in\mathbb{R}^{{I_1} \times {R_1}},\textbf{U}^{(2)}_0\in\mathbb{R}^{{I_2} \times {R_2}},\textbf{U}^{(3)}_0\in\mathbb{R}^{{I_3} \times {R_3}}\}$;\\
Obtain $\overline{\bm{\mathcal{X}}}$ by averaging all training patches;\\
Obtain $\bm{\mathcal{G}}^0$ by (7);\\
Flatten $\overline{\bm{\mathcal{X}}}$ along each mode and obtain three matricizations $\textbf{X}_{(1)},\textbf{X}_{(2)},\textbf{X}_{(3)}$;\\
\While {$\Vert\bm{\mathcal{G}}^{t-1}-\bm{\mathcal{G}}^{t}\Vert_F>0.01$,}
{
	Obtain $\textbf{U}^{(1),t}$ by (9);\\
	Obtain $\textbf{U}^{(2),t}$ by (10);\\
 	Obtain $\textbf{U}^{(3),t}$ by (11);\\
 	Obtain $\bm{\mathcal{G}}^t$ by (7), $t=t+1;$\\
}
\KwOut{The final $\textbf{U}^{(1),t}, \textbf{U}^{(2),t},\textbf{U}^{(3),t}$ are output as $\textbf{U}^{(1)}, \textbf{U}^{(2)},\textbf{U}^{(3)}$.}
\end{algorithm}

We use $\textbf{U}^{(1)}, \textbf{U}^{(2)},\textbf{U}^{(3)}$ as the mapping kernel of three mapping layers correspondingly. The input of mapping layer is multiplied by mapping kernel according to (4). The input patch first goes through mapping layers. For mapping layer 1, the patch is mapped to a new tensor with its first dimensionality reduced to $R_1$.
It means the original tensor can be looked as $I_2\times{I_3}$ vectors with $I_1$ features, and the mapping layer maps these $I_2\times{I_3}$ vectors into shorter vectors with only $R_1$ energy concentrated features. The mapping layer 2 and mapping layer 3 are sequentially connected, and each mapping layer applies corresponding mode multiplication by (4). The procedures are also shown in Fig. 2. The input of mapping layer 1 is HSI patch of size $I_1\times{I_2}\times{I_3}$, and the output of mapping layer 3 is a tensor of size $R_1\times{R_2}\times{R_3}$. Our mapping layers are used to extract the energy concentrated feature so that most energy of the input patch is still kept, which is because the weights in mapping kernel are actually the singular vectors of matricization.

\subsection{The Architecture of MCNN}
Since we apply three mapping layers in our network, the output of the mapping layer 3 is an energy concentrated tensor while the spatial-spectral structure of original patch is still reserved, which is beneficial for extracting the edge feature by convolution. As CNN has achieved human-level intelligence in several perception tasks\cite{lecun2015deep,mnih2015human}, we propose to use the convolutional layer to extract discriminative feature in our method. However, convolutional layer can hardly reduce the feature dimension, which will affect the scale of weights in the following fully connected layers. Therefore, in order to reduce its output dimension, we connect each convolutional layer with a pooling layer. Moreover, it is quite effective to extract spectral-spatial feature from our energy concentrated tensor. Thus, two convolutional layers are enough to extract features and compose the features as discriminative information. In this paper, we adopt max pooling in the pooling layers. Then, two fully connected layers are followed. The output of the last fully connected layer is a one-hot vector with length of class number. 

The architecture of the proposed MCNN includes three sections, and the first section is called mapping section, which is built for extracting the spectral-spatial feature. We take Indian Pines as an example to explain the proposed MCNN, the input of network is a HSI patch of size $13\times13\times200$, and three mapping layers have mapping kernels of sizes $13\times7$, $13\times7$ and $200\times40$, respectively. The input patch is mapped as a feature cube of size $7\times7\times40$. The energy in the HSI patch is concentrated and most significant dimensions are reserved. The second section is a convolutional section, which includes two convolutional layers and two max pooling layers. This section is mainly built for extracting the edge features and composing them as abstract features for classification. The max pooling layer can reduce the feature dimension and it usually follows the convolutional layer. The input of the first convolutional layer is $7\times7\times40$. We use a  $5\times5\times10$ tensor as a convolutional kernel in this layer, and the stride of (1,1,5) is applied to subsampling the input. There are 64 kernels in this convolutional layer, and the output is a fourth-order tensor of size  $5\times5\times54\times64$. A max pooling layer with pooling size of  $3\times3\times5$ and stride of (1,1,2) is followed. We convolve the output of pooling layer with another 64 kernels of sizes $5\times5\times10$ as the second convolutional layer. A max pooling layer with pooling size of $3\times3\times5$ and stride of (1,1,2) is followed as well. These four layers construct the convolutional section. The fully connected section of two fully connected layers is mainly used to classify the obtained the spectral-spatial features. Fig. 2 shows the architecture of our method.
\begin{figure*}[htpb]
\centering
\includegraphics[width=0.95\textwidth]{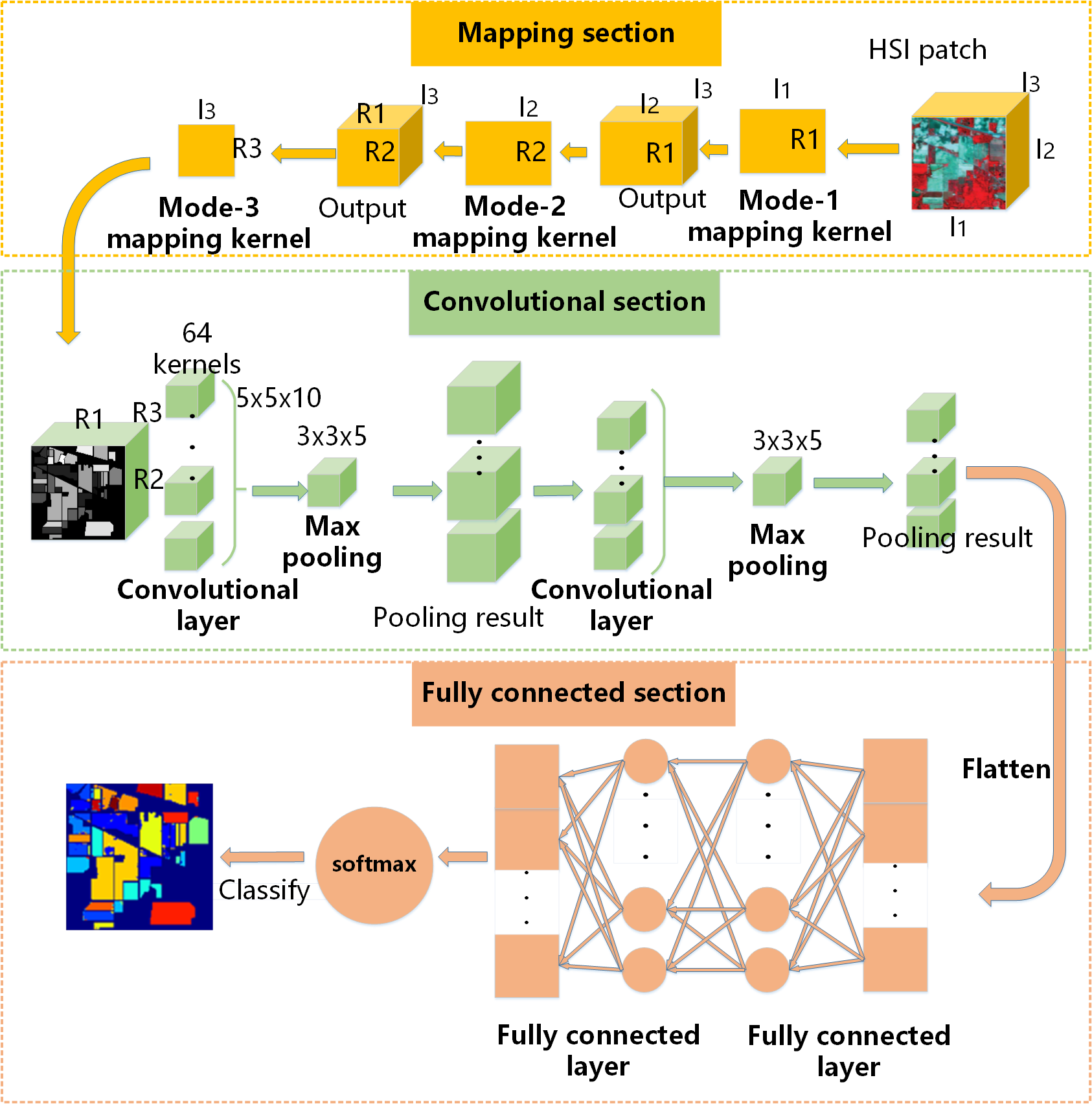}
\caption{The architecture of our proposed MCNN.}
\label{fig3}
\end{figure*}

The main difference between traditional CNN and our MCNN is the mapping layer. The main purposes of adopting these mapping layers are to more effectively extract the spectral-spatial feature and more efficiently reduce the redundancy. Therefore, multilinear algebra is applied in mapping layer. We obtain our mapping kernels by optimizing the TD problem, and these kernels can well keep the spectral-spatial structure and reduce the unnecessary dimension.
\subsection{The 3-D Convolutional Kernel}
Many CNN architectures have to deal with the large spectral redundancy, and the widely used PCA is applied to reduce the spectral dimension. The first three principal components (PCs) are usually extracted from HSI by PCA. However, these methods may lose plenty of spectral information. Considering this, we adopt the 3-D convolutional kernel, which is firstly developed by Chen\cite{chen2016deep} to effectively extract the spectral-spatial feature in HSI. 

The 3-D convolutional kernel performs convolution not only on spatial direction, but also on spectral direction. The output of one kernel is a third-order tensor.
The output at position $(x,y,z)$ can be expressed as follows:
\begin{equation}
v_{x,y,z}=f(\sum^{T}_{t=0}\sum^{P}_{p=-P}\sum^{Q}_{q=-Q}
\sum^{R}_{r=-R}{w_{t,p,q,r}\alpha_{(x+p),(y+q),(z+r)}}+b_{x,y,z}),
\end{equation}
where $(x,y,z)$ is the position in input patch, $(2P+1), (2Q+1), (2R+1)$ are the corresponding sizes of convolutional kernel, $T$ is the number of channels (convolutional kernels), $w$ and $b$ are weights in convolutional kernel. The weights $w$ and $b$ of 3-D convolutional kernel are all randomly initialized by zero mean Gaussian distribution. They are iteratively updated through the back propagation procedure. $f(\cdot)$ is nonlinear function to improve the nonlinear mapping capability.
Since the output of convolutional layer is a third-order tensor, the pooling layer is also a 3-D pooling layer, and the strides in convolutional layers and pooling layers are also on three directions.
\begin{figure}[htpb]
\centering
\includegraphics[width=0.6\columnwidth]{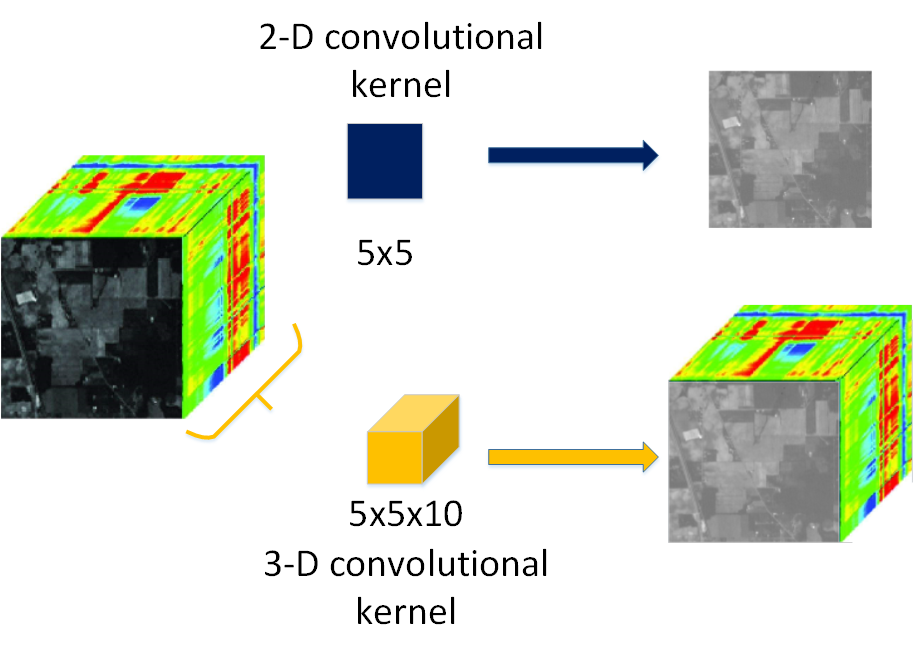}
\caption{The comparison of two kinds of convolutional kernels.}
\label{fig4}
\end{figure}

We show the comparison of two kinds of convolutional kernels in Fig. 3. The advantage of 3-D convolutional kernel is that it can extract spectral-spatial information simultaneously. For HSI classification, the structure information exists in the HSI patch, and 2-D convolution may not extract it completely. Thus, applying 3-D convolution is needed.
Meanwhile, 3-D convolution involves less parameters than other deep learning-based methods at the same scale. Thus, it is more appropriate for problems with limited training samples, like HSI classification.
In this paper, the 3-D convolutional kernel is of size $5\times5\times10$. 
\\
\section{The Experimental Results and Analysis}
In this section, we introduce three commonly used HSI datasets, some experimental details and results.We first validate the hyperparameters of our MCNN. Then, we compare our proposed MCNN with several state-of-the-art methods, and we also analyze the computational time of our network against other deep learning methods. Finally, we test the efficiency of our mapping layers. All of these experiments are tested on a computer of i7 7700 CPU @ 3.60GHz 24GB of RAM and GeForce GTX1080Ti GPU.
\subsection{The HSI Datasets}
In this section, we first introduce three HSI datasets: Indian Pines, Pavia University and Salinas. Then we give our specified hyperparameters of our network, and we compare our network with several state-of-the-art HSI classification methods including 3 deep learning methods, which are spectral-spatial residual network (SSRN) \cite{zhong2018}, deformable convolutional neural network (DHCNet) \cite{zhu2018deformable} and 3DCNN \cite{chen2016deep}\cite{li2017spectral}, and 2 non-deep-learning methods: edge preserve filtering (EPF) \cite{kang2014spectral} and support vector machine (SVM). There are two 3DCNNs in our experiment: 3DCNN$_1$ has the same convolutional section with our MCNN (only 2 convolutional layers), 3DCNN$_2$ has six convolutional layers. All these methods are evaluated by classification metrics: overall accuracy (OA), average accuracy (AA) and kappa coefficient ($\kappa$). In all these examinations, we take twenty percent of HSI patches as training samples. Ten percent of HSI patches are picked randomly from the rest HSI dataset as validation set, which do not include the training set. And the rest seventy percent of HSI patches are test samples.

The Indian Pines (IP) dataset is collected by Airborne Visible/Infrared Imaging Spectrometer (AVIRIS) in Northwestern Indiana, which contains 16 vegetation classes. There are $145\times145$ pixels with resolution of 20 m. There are 200 spectral bands in a range from 400 to 2500nm, 24 bands corrupted by water absorption effect and noise are discarded. The false-color composite of Indian Pines image and its ground truth map are shown in Fig. 4.
\begin{figure}[h]
\raggedright
\subfigure[]{

\includegraphics[width=0.3\columnwidth]{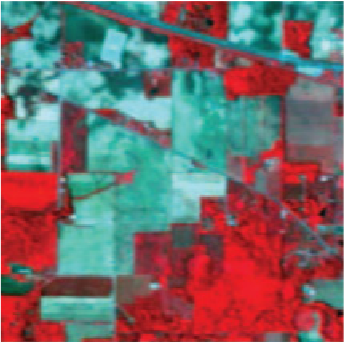}}
\subfigure[]{

\includegraphics[width=0.3\columnwidth]{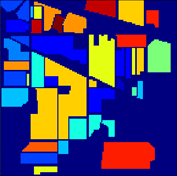}}
\subfigure[]{

\includegraphics[width=0.3\columnwidth]{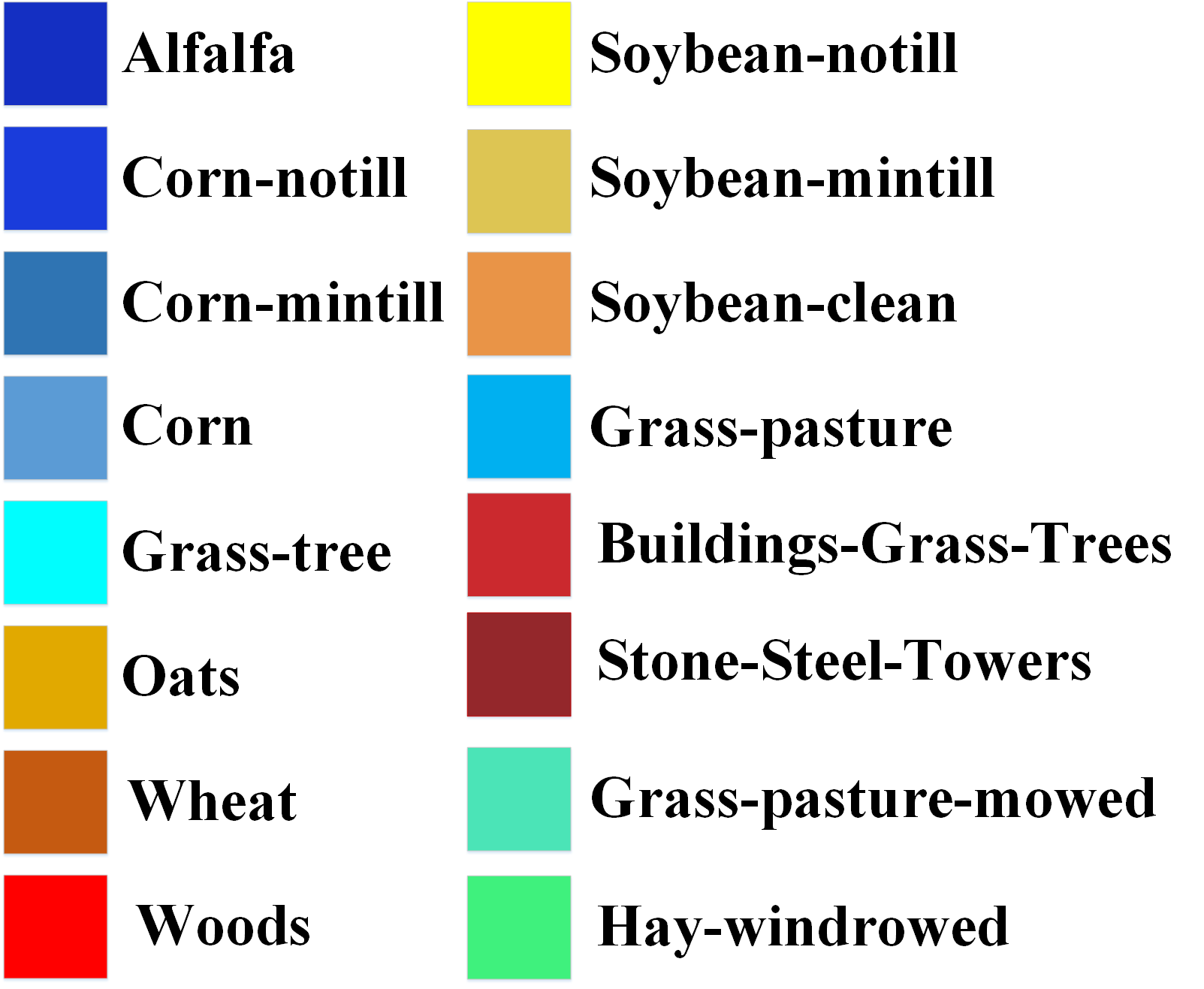}}
\caption{(a) Three-band color composite of the Indian Pines image; (b) ground truth of Indian Pines; (c) color map.}
\label{fig5}
\end{figure}

The University of Pavia (UP) dataset is acquired by Reflective Optics System Imaging Spectrometer in Northern Italy in 2001, which contains nine classes. There are $610\times340$ pixels with resolution of 1.3 m per pixel. There are 115 spectral bands in a range of 430 to 860nm, and 12 most noisy bands are removed before experiment. The false-color composite of the University of Pavia image and its corresponding ground truth map are shown in Fig. 5.
\begin{figure}[h]
\centering
\subfigure[]{

\includegraphics[width=0.3\columnwidth]{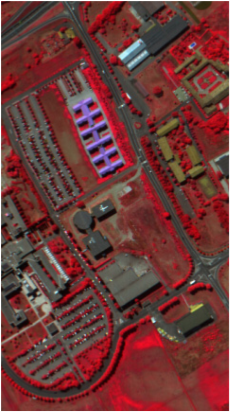}}
\subfigure[]{

\includegraphics[width=0.3\columnwidth]{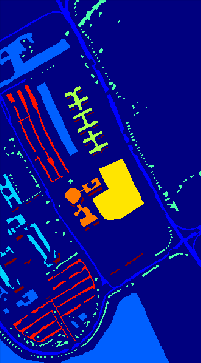}}
\subfigure[]{

\includegraphics[width=0.15\columnwidth]{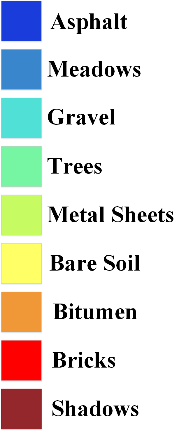}}
\caption{(a) Three-band color composite of the UP image; (b) ground truth of UP; (c) color map.}
\label{fig6}
\end{figure}

The Salinas dataset is collected by AVIRIS in Florida, which contains 16 classes. There are $512\times217$ pixels with resolution of 3.7 m. There are 224 spectral bands in a range from 400 to 2500nm, and all bands are reserved. The false-color composite of the Salinas image and its corresponding ground truth map are shown in Fig. 6.
\begin{figure}[h]
\centering
\subfigure[]{

\includegraphics[width=0.25\columnwidth]{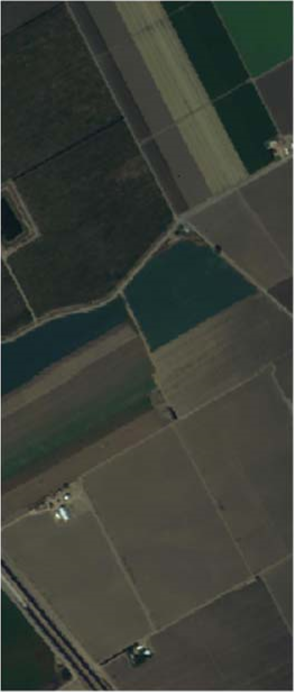}}
\subfigure[]{

\includegraphics[width=0.25\columnwidth]{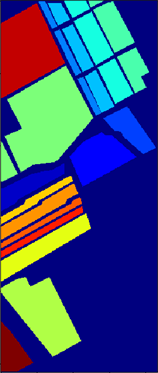}}
\subfigure[]{

\includegraphics[width=0.33\columnwidth]{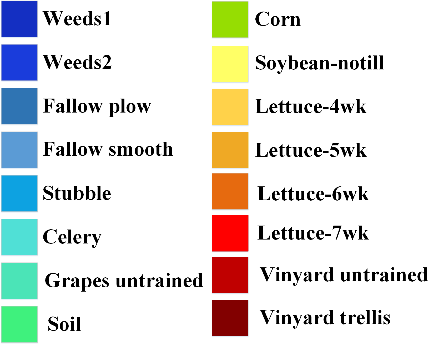}}
\caption{(a) Three-band color composite of the Salinas image; (b) ground truth of Salinas; (c) color map.}
\label{fig3}
\end{figure}

The hyperparameters of comparative methods are set according to the published paper. The sizes of input patch of 3DCNN$_1$, 3DCNN$_2$, SSRN and our MCNN are $13\times 13\times band$, where $band$ is the number of spectral bands. Only DHCNet is $25\times 25\times band$, because this method requires larger spatial size. Thanks to those authors' efforts, SSRN\footnote{https://github.com/zilongzhong/SSRN}, DHCNet\footnote{https://github.com/OrdianryCore/DHCNet} and EPF\footnote{http://xudongkang.weebly.com/} are open sourced. The SVM algorithm is implemented in the LIBSVM\footnote{https://www.csie.ntu.edu.tw/~cjlin/libsvm/} library \cite{chang2011libsvm}. Our code will be  open sourced in Github\footnote{https://github.com/dashaqi/MCNN}.

\subsection{The Validation of Hyperparameters}

In this subsection, there are two kinds of hyperparameters needing validation, which are (1) the hyperparameters of convolutional networks, including learning rate, training epoch and the batch size; (2) the hyperparameters of mapping layers, including output sizes of mapping layers: $R_1$, $R_2$, $R_3$. We will first validate the hyperparameters of convolutional network, and then validate hyperparameters of our mapping layers based on the model of the validated hyperparameters. There are two reasons we can validate these two kinds of parameters separately. Firstly, our mapping layers are a kind of preprocessing method for the input patch. The mapping layers are not updated by back propagation, thus, they are not influenced by learning rate. Secondly, mapping layers extract spectral-spatial feature from each single input patch, which means there is no need to validate batch size with the hyperparameters of mapping layers.

For the convolutional blocks in our MCNN architecture, there are three hyperparameters that can control the training process and the classification performance. They are learning rate, training epoch and the batch size in convolutional layer. We use Adam optimizer in our work, $R_1$, $R_2$, $R_3$ are specified as 7, 7, 40 temporally for three datasets in accordance with our experience. The learning rate and batch size are validated together to find the best combination. Speciﬁcally, inappropriate learning rate settings will lead to divergence or slow convergence. Thus, we give ranges of batch size and learning rate as: (20, 30, 40), (0.01, 0.003, 0.001, 0.0003, 0.0001), respectively. The best combination of batch size and learning rate is searched among these two ranges. The result of trained model is output every 10 epochs, and we take the epoch number of the best validation results as the validated epoch number. 

\begin{figure}[htpb]
\centering
\subfigure[]{
\includegraphics[width=0.8\columnwidth]{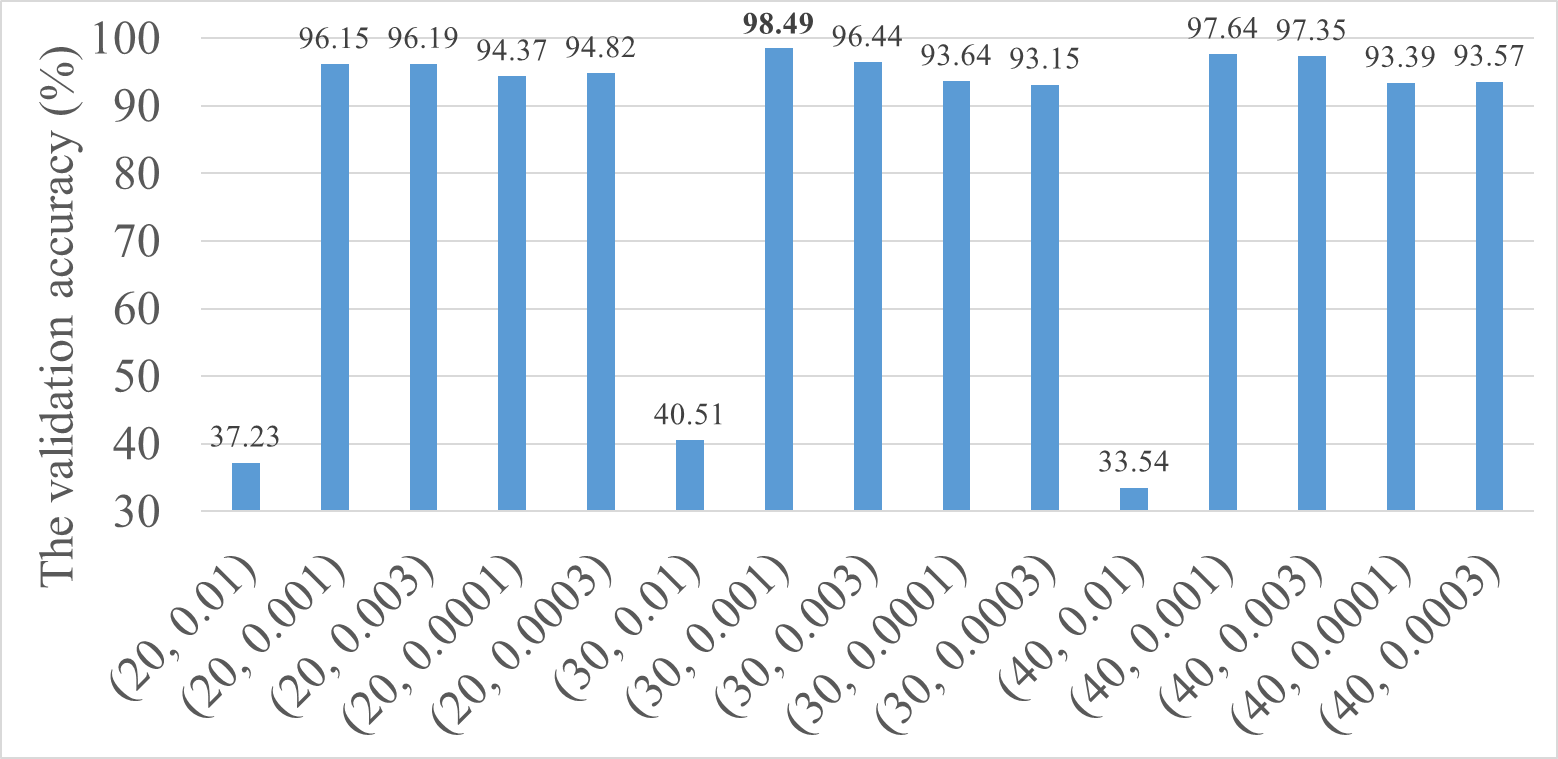}}
\subfigure[]{
\includegraphics[width=0.8\columnwidth]{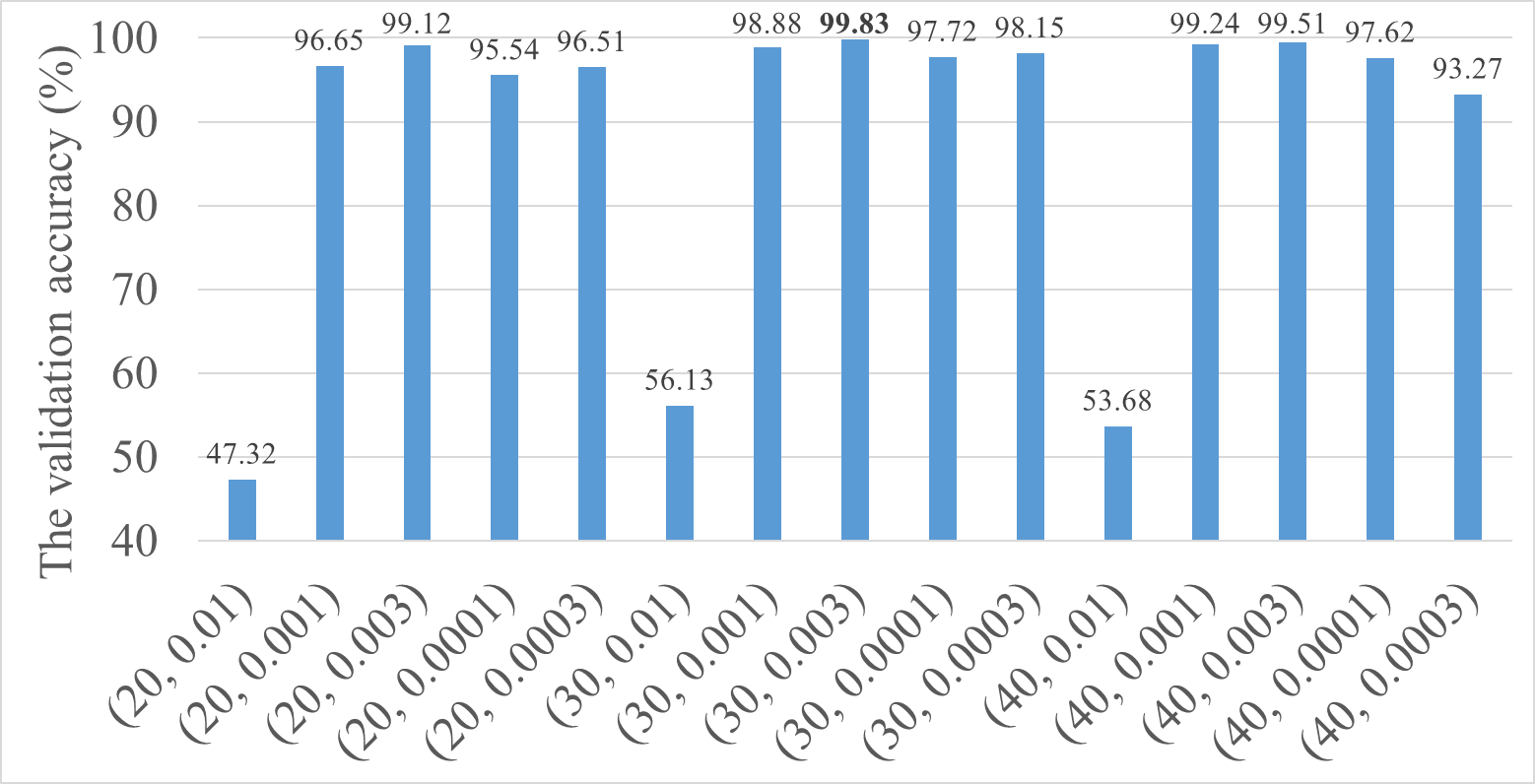}}
\subfigure[]{
\includegraphics[width=0.8\columnwidth]{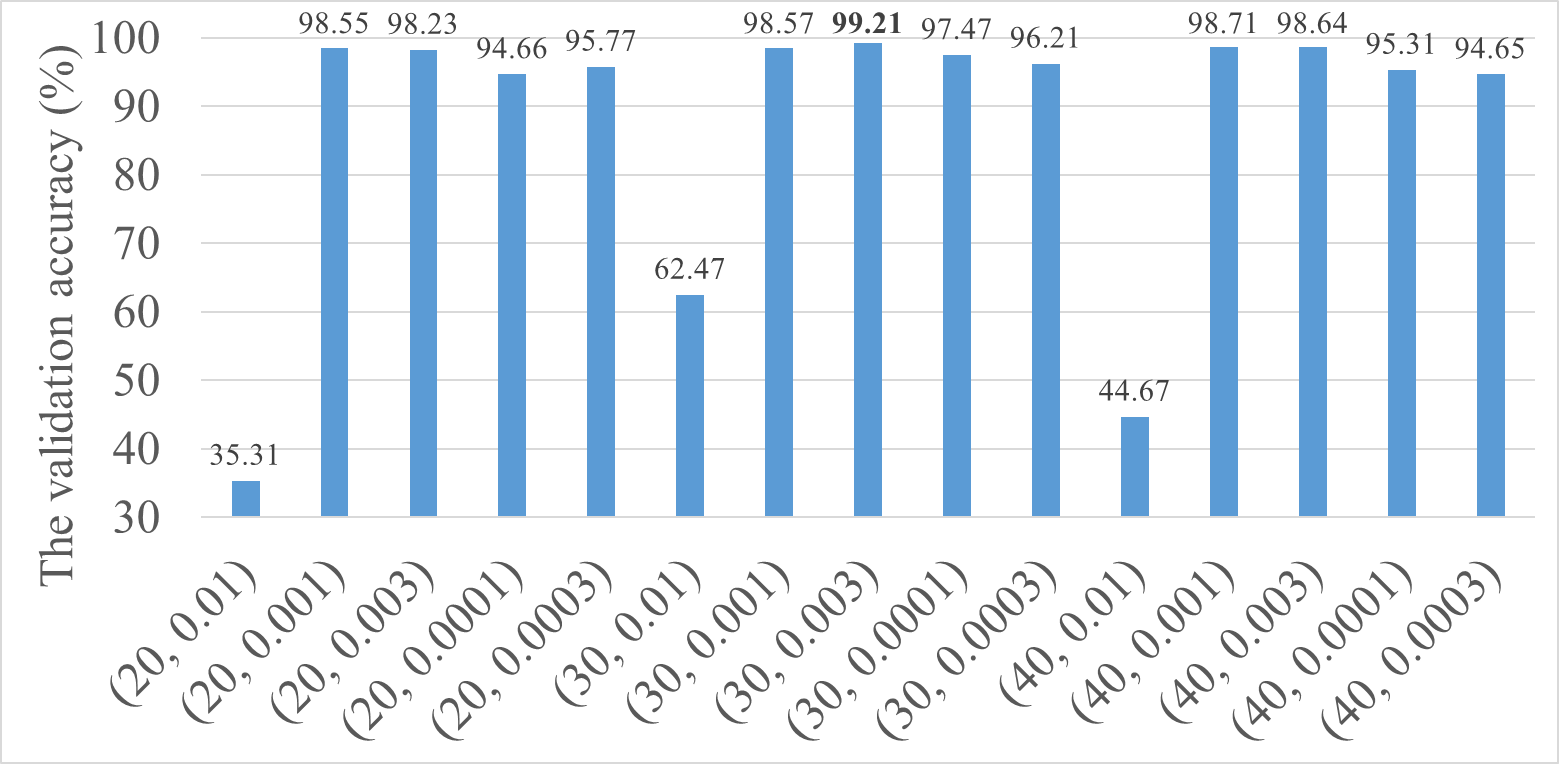}}
\caption{The validation results of 15 models on three datasets, each model is the combination of batch size and learning rate, (a) Indian Pines, (b) University of Pavia, (c) Salinas.}
\label{fig7}
\end{figure}

All the validation accuracies in Fig. 7 are achieved on the best epoch number. It can be seen from Fig. 7 (a) that the model with batch size of 30 and learning rate of 0.001 performs best on the validation set of Indian Pines (IP). To find out the best epoch number of model (30, 0.001), we show the validation accuracy varying with epoch on the best model (30, 0.001) of IP in Fig. 8 (a). When the model is trained 30 epochs, it can obtain the best performance on validation set. Thus, the most suitable hyperparameters of our MCNN on Inidan Pines is that learning rate is specified as 0.001, batch size specified as 30 and training epoch specified as 30.

\begin{figure}[htpb]
\centering
\subfigure[]{
\includegraphics[width=0.4\columnwidth]{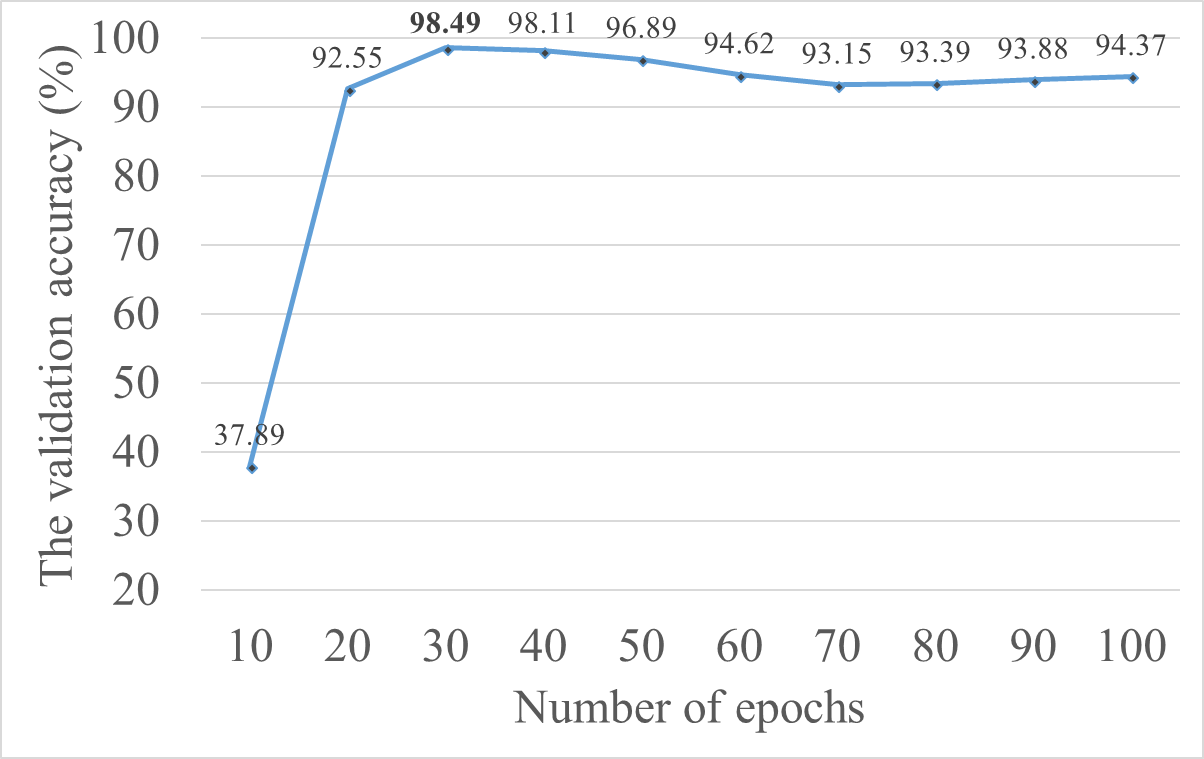}}
\subfigure[]{
\includegraphics[width=0.4\columnwidth]{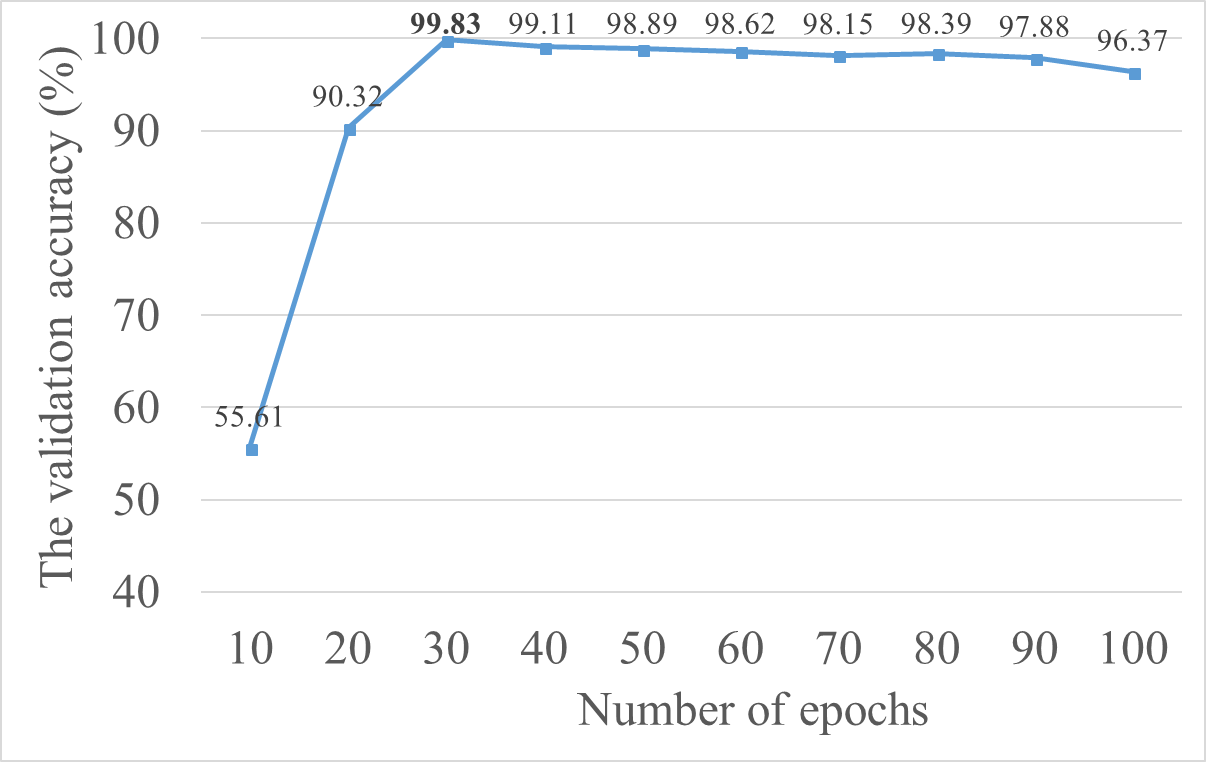}}

\subfigure[]{
\includegraphics[width=0.4\columnwidth]{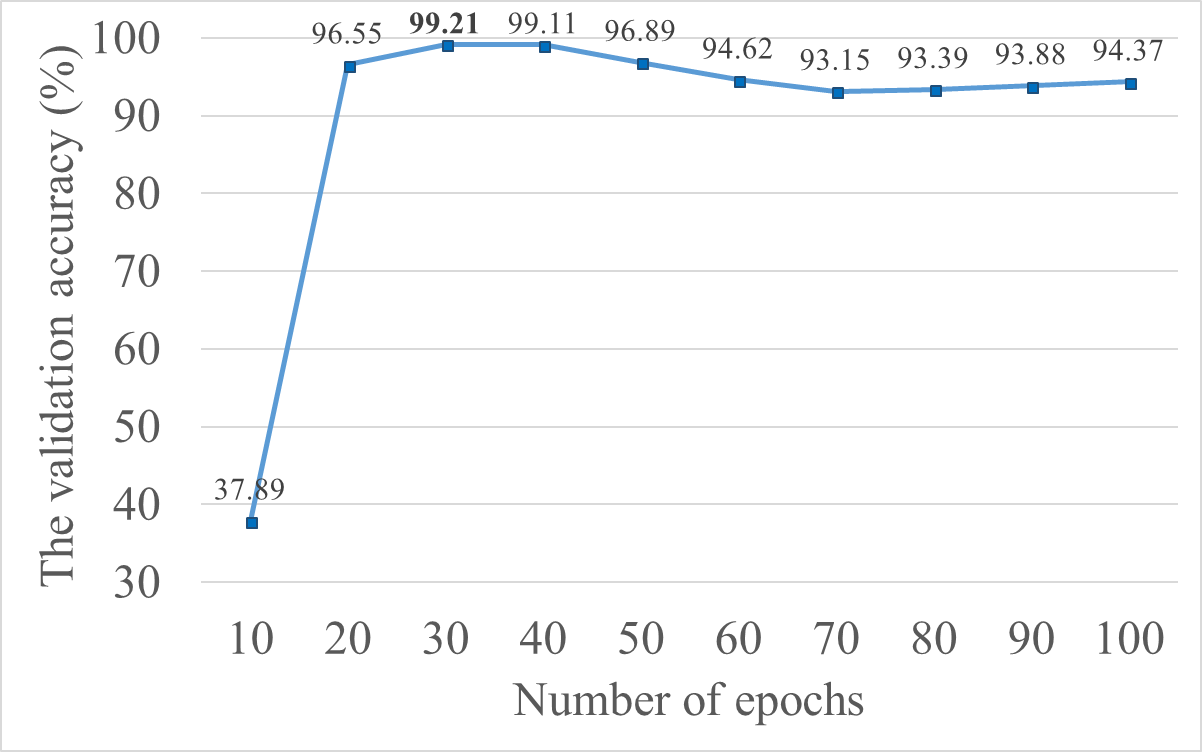}}
\caption{The validation accuracy varies with the epoch number on the best model (batch size, learning rate): (a) the validation results of Indian Pines on the model (30, 0.001); (b) the validation results of University of Pavia on the model (30, 0.003); (c) the validation results of Salinas on the model (30, 0.003).}
\label{fig8}
\end{figure}

We also test the 15 models on University of Pavia (UP) and Salinas in the same way. The validation results are also given in Fig. 7 (b), Fig. 8 (b) and Fig. 7 (c), Fig. 8 (c). The results show that the best hyperparameters for University of Pavia and Salinas are the same. For University of Pavia and Salinas, we set learning rate as 0.003, batch size as 30 and training epoch as 30. 

As the hyperparameters of convolutional blocks are validated, then, we will validate the hyperparameters of $R_1$, $R_2$, $R_3$ in our mapping layers. Since the independence of mapping layers, we can validate the hyperparameters of mapping layers and convolutional blocks separately.

The mapping layers are designed to extract the spectral-spatial feature. The input patch of our MCNN is a third-order tensor, we use three mapping layers to extract the features of each order of the patch. Each mapping layer has a hyperparameter $R_n, n=1,2,3$. These three hyperparameters can determine the sizes of extracted spectral-spatial feature, which can further determine the quality of features and computational consumption of MCNN. To find out the best combination of  $R_1$, $R_2$, $R_3$, we validate them on the validation dataset. The hyperparameters of batch size and learning rate are set as (30, 0.001) for IP, the hyperparameters of other two datasets are set as (30, 0.003), (30, 0.003) for UP and Salinas respectively.

Considering our input is a third-order HSI patch,  where $R_1$ and $R_2$ are spatial sizes, $R_3$ is spectral size, we select the same value for $R_1$ and $R_2$. We validated $R_1$, $R_2$, $R_3$ on three datasets: IP, UP and Salinas. The validated combination of these hyperparameters and their classification accuracies are listed below.
\begin{table*}[htpb]
   \centering
   \caption{The validation accuracy (\%) varies with hyperparameters $R_1$; $R_2$; $R_3$ on Indian Pines.}
   \label{tab2}
   \begin{tabular}{p{0.1\linewidth}<{\centering}p{0.08\linewidth}<{\centering}p{0.08\linewidth}<{\centering}p{0.08\linewidth}<{\centering}p{0.08\linewidth}<{\centering}p{0.08\linewidth}<{\centering}p{0.08\linewidth}<{\centering}}
   \toprule
&&\multicolumn{5}{c}{$R_3$}\\
\cline{3-7}
&&20&40&60&100&140\\
\cline{3-7}
\multirow{4}{*}{$R_1, R_2$}&11&95.05&98.06&95.14&96.56&95.60\\
&9&96.27&97.94&96.38&96.98&92.35\\
&7&93.51&\textbf{98.49}&97.18&96.39&94.56\\
&5&92.34&95.14&94.64&93.58&93.54\\
    \bottomrule
   \end{tabular}
\end{table*}	

\begin{table*}[htpb]
   \centering
   \caption{The validation accuracy (\%) varies with hyperparameters $R_1$, $R_2$, $R_3$ on University of Pavia.}
   \label{tab2}
   \begin{tabular}{p{0.1\linewidth}<{\centering}p{0.08\linewidth}<{\centering}p{0.08\linewidth}<{\centering}p{0.08\linewidth}<{\centering}p{0.08\linewidth}<{\centering}p{0.08\linewidth}<{\centering}p{0.08\linewidth}<{\centering}}
   \toprule
&&\multicolumn{5}{c}{$R_3$}\\
\cline{3-7}
&&20&30&40&50&60\\
\cline{3-7}
\multirow{4}{*}{$R_1, R_2$}&11&99.01&97.06&99.64&99.65&99.68\\
&9&99.62&98.94&99.77&99.75&99.71\\
&7&\textbf{99.88}&99.12&99.83&99.73&99.60\\
&5&99.05&99.14&99.70&99.70&99.71\\
    \bottomrule
   \end{tabular}
\end{table*}

\begin{table*}[htpb]
   \centering
   \caption{The validation accuracy (\%) varies with hyperparameters $R_1$, $R_2$, $R_3$ on Salinas.}
   \label{tab2}
   \begin{tabular}{p{0.1\linewidth}<{\centering}p{0.08\linewidth}<{\centering}p{0.08\linewidth}<{\centering}p{0.08\linewidth}<{\centering}p{0.08\linewidth}<{\centering}p{0.08\linewidth}<{\centering}p{0.08\linewidth}<{\centering}}
   \toprule
&&\multicolumn{5}{c}{$R_3$}\\
\cline{3-7}
&&20&40&60&100&140\\
\cline{3-7}
\multirow{4}{*}{$R_1, R_2$}&11&97.46&98.51&98.44&98.64&97.68\\
&9&98.36&98.80&98.42&98.41&99.15\\
&7&98.40	&\textbf{99.21}&98.45&96.39&98.89\\
&5&96.93&97.66&96.68&98.61&96.47\\
    \bottomrule
   \end{tabular}
\end{table*}	

The validation results show that the best hyperparameters of our mapping layers are different for three datasets. According to the results in Table II, Table III and Table IV, we specify ($R_1$, $R_2$, $R_3$) as (7, 7, 40), (7, 7, 20) and (7, 7, 40) for IP, UP and Salinas, respectively. The results also reveal that our mapping layers are able to extract better spectral-spatial features of small sizes, which further reduce computational cost and achieve better classification results.

\subsection{Classification Performance}
We test our methods with six comparative methods of SSRN, DHCNet, 3DCNN$_1$, 3DCNN$_2$, EPF and SVM on Indian Pines, University of Pavia and Salinas. Tables V-VII show the accuracy reports on three datasets. Fig. 9-11 show the classification maps obtained by different methods associated with the corresponding OA scores. 

From these figures, it can be seen that the classification obtained by SVM is not satisfactory since some noisy estimations are still visible. Among the deep learning methods, 3DCNN$_2$ performs better than 3DCNN$_1$ for the reason that the former has more convolutional layers than the latter. Although the classification performance decreases when more convolutional layers are deployed in CNN, two convolutional layers in 3DCNN$_1$ are clearly not enough to extract feature from an input patch of sizes $13\times13\times200$. We specify the framework of 3DCNN$_1$ to compare with our MCNN as they both apply two convolutional layers. Our MCNN obtains the best OA scores, especially our MCNN increases the OA compared with 3DCNN$_1$ method by about $2\%-7\%$. These two methods have the same layers and architectures except our proposed mapping layers, which verify the effectiveness of our mapping layers. 

Our mapping layer can extract better spectral-spatial feature, and the feature concentrates the energy of the patch. Thus, mapping layers can reduce the redundancy and accelerate the following convolution options. As the mapping layer is designed for HSI (the kernel is not updated by back propagation), it can obtain more effective features much faster than convolutional layers. Tables V-VII report the OAs, AAs, kappa coefficients and the classification accuracies of all classes for HSI classification. In three cases, our MCNN achieves the highest classification accuracy.
\begin{figure}[htpb]
\centering
\subfigure[]{
\includegraphics[width=0.2\columnwidth]{Indian_Pines_groundtruth.png}}
\subfigure[]{
\includegraphics[width=0.2\columnwidth]{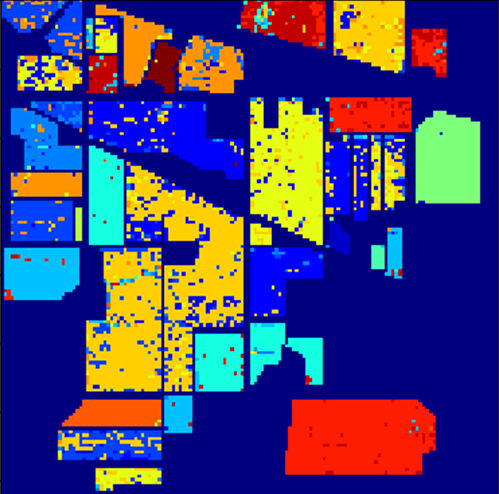}}
\subfigure[]{
\includegraphics[width=0.2\columnwidth]{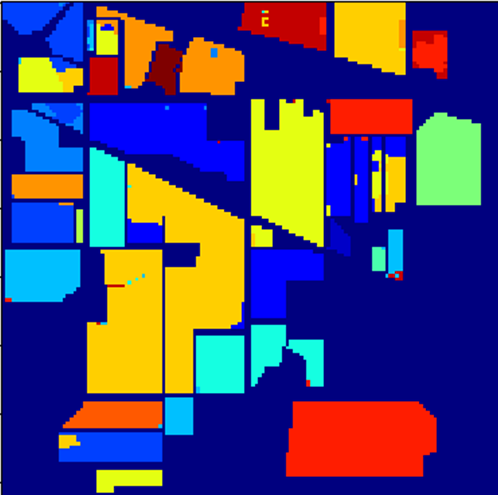}}
\subfigure[]{
\includegraphics[width=0.2\columnwidth]{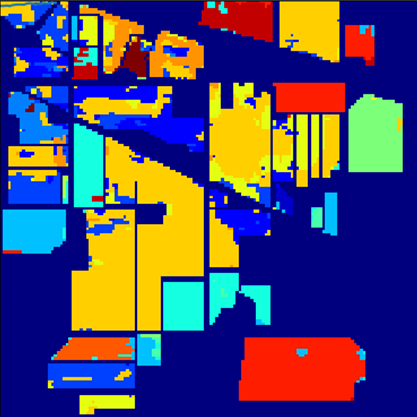}}

\subfigure[]{
\includegraphics[width=0.2\columnwidth]{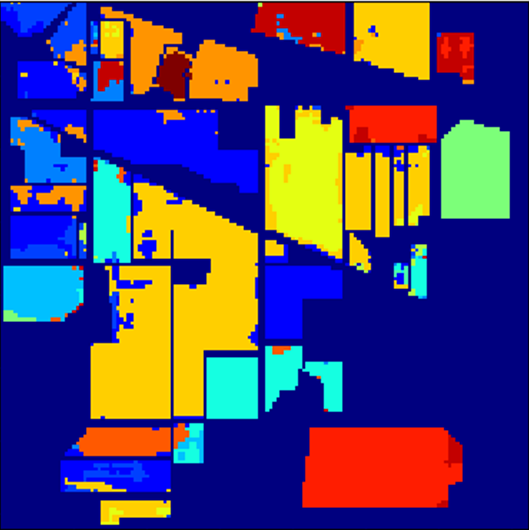}}
\subfigure[]{
\includegraphics[width=0.2\columnwidth]{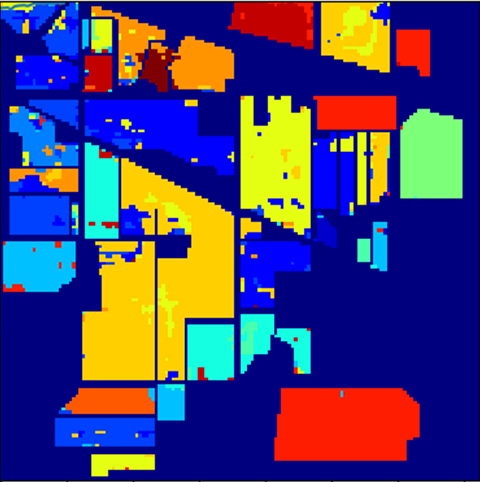}}
\subfigure[]{
\includegraphics[width=0.2\columnwidth]{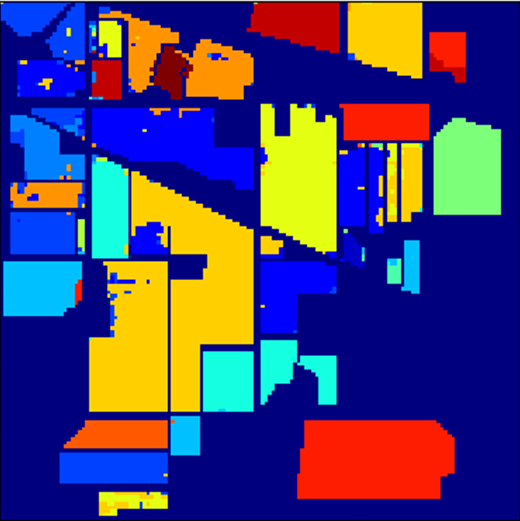}}
\subfigure[]{
\includegraphics[width=0.2\columnwidth]{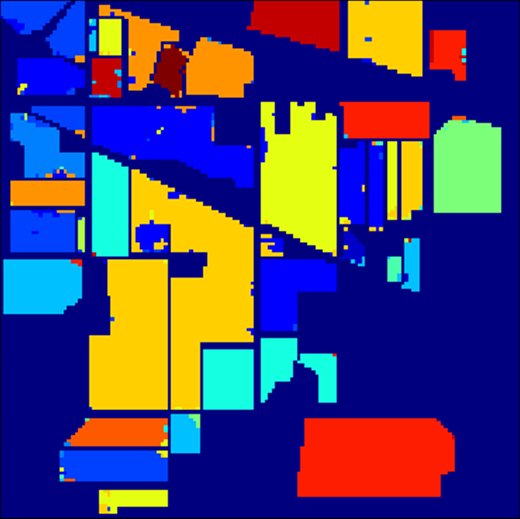}}
\caption{Classification results of all methods for Indian Pines, (a) ground truth; (b)-(h) classification results of SVM, EPF, 3DCNN$_1$,3DCNN$_2$, DHCNet, SSRN, and MCNN.}
\label{fig7}
\end{figure}
\begin{figure}[htpb]
\centering
\subfigure[]{
\includegraphics[width=0.15\columnwidth]{UP_groundtruth.png}}
\subfigure[]{
\includegraphics[width=0.15\columnwidth]{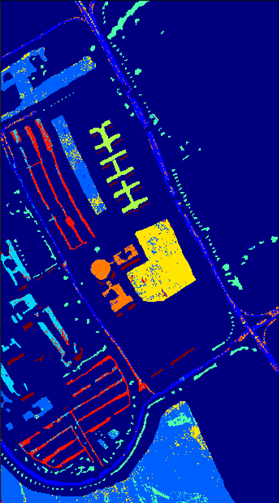}}
\subfigure[]{
\includegraphics[width=0.15\columnwidth]{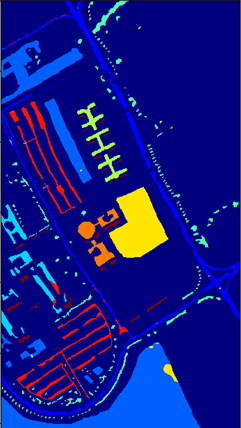}}
\subfigure[]{
\includegraphics[width=0.15\columnwidth]{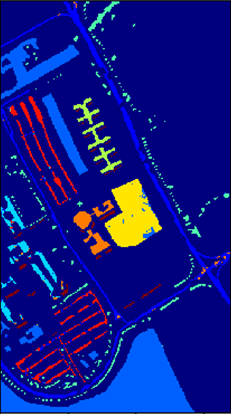}}

\subfigure[]{
\includegraphics[width=0.15\columnwidth]{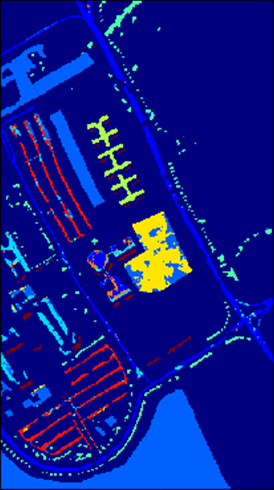}}
\subfigure[]{
\includegraphics[width=0.15\columnwidth]{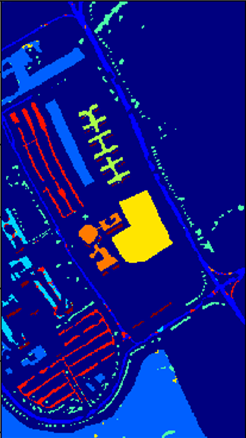}}
\subfigure[]{
\includegraphics[width=0.15\columnwidth]{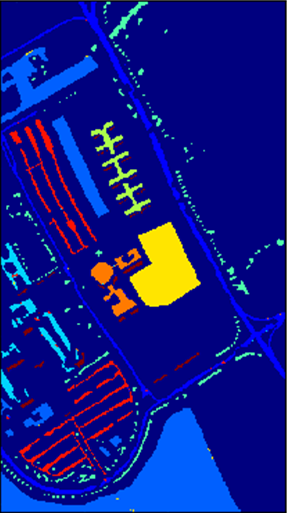}}
\subfigure[]{
\includegraphics[width=0.15\columnwidth]{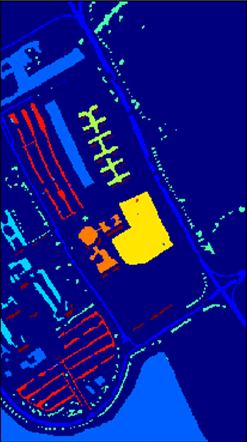}}
\caption{Classification results of all methods for University of Pavia, (a) ground truth; (b)-(h) classification results of SVM, EPF, 3DCNN$_1$,3DCNN$_2$, DHCNet, SSRN, and MCNN.}
\label{fig7}
\end{figure}
\begin{figure}[htpb]
\centering
\subfigure[]{
\includegraphics[width=0.1\columnwidth]{Salinas_groundtruth.png}}
\subfigure[]{
\includegraphics[width=0.1\columnwidth]{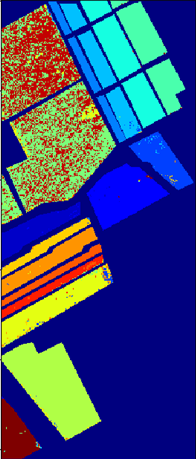}}
\subfigure[]{
\includegraphics[width=0.1\columnwidth]{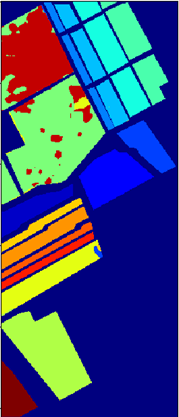}}
\subfigure[]{
\includegraphics[width=0.1\columnwidth]{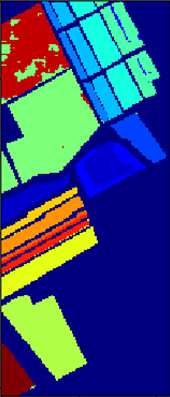}}
\subfigure[]{
\includegraphics[width=0.1\columnwidth]{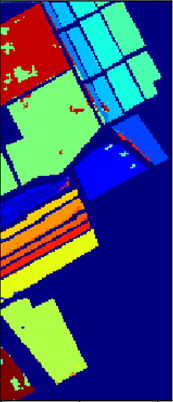}}
\subfigure[]{
\includegraphics[width=0.1\columnwidth]{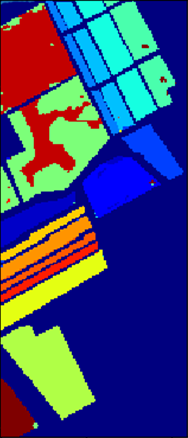}}
\subfigure[]{
\includegraphics[width=0.1\columnwidth]{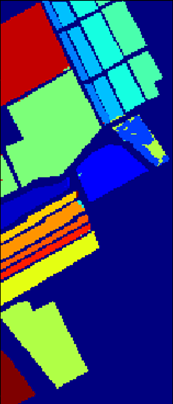}}
\subfigure[]{
\includegraphics[width=0.1\columnwidth]{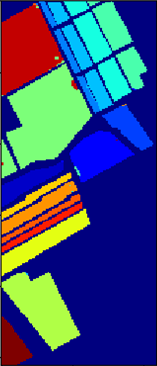}}
\caption{Classification results of all methods for Salinas, (a) ground truth; (b)-(h) classification results of SVM, EPF, 3DCNN$_1$,3DCNN$_2$, DHCNet, SSRN, and MCNN.}
\label{fig7}
\end{figure}
\begin{table*}[htpb]
   \centering
   \caption{The classification accuracies (\%) of our MCNN and six comparative methods tested on Indian Pines}
   \label{tab2}
   \begin{tabular}{p{0.10\linewidth}<{\centering}p{0.09\linewidth}<{\centering}p{0.09\linewidth}<{\centering}p{0.09\linewidth}<{\centering}p{0.09\linewidth}<{\centering}p{0.09\linewidth}<{\centering}p{0.09\linewidth}<{\centering}p{0.09\linewidth}<{\centering}}
   \toprule
Class&SVM&EPF&3DCNN$_1$&3DCNN$_2$&DHCNet&SSRN&MCNN\\  
    \midrule
Alfalfa&$85.6\pm1.2$&$98.3\pm0.3$&$80.1\pm1.2$&$95.4\pm0.6$&$75.7\pm0.3$&$97.7\pm0.3$&$98.3\pm0.7$\\
Corn-N&$80.1\pm1.0$&$93.4\pm0.2$&$75.5\pm2.2$&$85.6\pm0.5$&$92.3\pm0.2$&$97.3\pm2.2$&$96.2\pm0.6$\\
Corn-M&$70.4\pm2.5$&$97.3\pm0.4$&$85.2\pm1.2$&$89.2\pm0.3$&$92.2\pm1.5$&$99.1\pm0.4$&$95.3\pm1.8$\\
Corn&$80.1\pm1.8$&$85.2\pm2.8$&$86.3\pm1.2$&$84.6\pm0.7$&$85.6\pm1.8$&$97.5\pm1.7$&$94.2\pm1.3$\\
Grass-M&$88.1\pm1.1$&$98.4\pm0.3$&$91.2\pm0.8$&$94.3\pm0.5$&$82.7\pm2.8$&$97.2\pm0.8$&$98.2\pm0.2$\\
Grass-T&$94.4\pm0.3$&$96.7\pm0.2$&$98.1\pm0.2$&$98.3\pm0.2$&$92.3\pm2.0$&$98.6\pm1.2$&$98.7\pm0.7$\\
Grass-P-M&$88.1\pm1.1$&$95.4\pm2.4$&$98.3\pm0.7$&$80.2\pm2.2$&$71.2\pm3.2$&$95.5\pm0.5$&$96.4\pm1.3$\\
  Hay-W&$98.5\pm1.2$&$100\pm0$&$96.2\pm0.8$&$99.2\pm0.2$&$86.3\pm1.2$&$94.8\pm0.8$&$100\pm0$\\
  Oats&$80.3\pm3.2$&$100\pm0$&$98.2\pm1.1$&$100\pm0$&$78.2\pm4.3$&$98.6\pm0.2$&$92.2\pm1.7$\\
  Soybean-N&$76.5\pm3.8$&$88.2\pm2.4$&$90.2\pm1.8$&$85.2\pm1.3$&$80.8\pm3.7$&$99.5\pm0.3$&$97.8\pm0.5$\\
  Soybean-M&$82.5\pm1.5$&$92.2\pm0.4$&$85.1\pm1.2$&$87.2\pm1.0$&$92.3\pm2.2$&$95.8\pm0.8$&$99.6\pm0.2$\\
  Soybean-C&$78.5\pm2.6$&$94.6\pm0.2$&$92.3\pm0.4$&$83.8\pm1.2$&$86.2\pm4.2$&$92.1\pm1.3$&$97.2\pm1.2$\\
  Wheat&$93.5\pm0.3$&$99.6\pm0.3$&$99.4\pm0.2$&$98.9\pm0.4$&$91.3\pm2.4$&$100\pm0$&$100\pm0$\\
  Woods&$94.2\pm0.4$&$99.3\pm0.1$&$98.2\pm0.4$&$99.2\pm0.3$&$97.2\pm1.1$&$99.2\pm0.2$&$99.7\pm0.2$\\
  Buildings-G-T-D	&$75.2\pm3.1$&$88.1\pm2.4$&$93.1\pm0.6$&$97.2\pm1.2$&$86.3\pm1.5$&$98.5\pm0.6$&$98.8\pm0.3$\\
  Stone-S-T&$88.0\pm2.5$&$89.0\pm1.2$&$92.1\pm0.8$&$92.5\pm1.5$&$76.3\pm4.5$&$98.6\pm0.6$&$95.8\pm1.5$\\
  OA&$83.7\pm0.3$&$94.3\pm0.3$&$88.9\pm0.3$&$90.7\pm0.5$&$90.1\pm1.1$&$97.4\pm0.3$&$98.3\pm0.2$\\
  AA&$84.6\pm0.4$&$94.7\pm0.4$&$91.2\pm0.4$&$91.9\pm0.5$&$85.4\pm0.6$&$97.5\pm0.3$&$97.4\pm0.4$\\
  $\kappa\times100$&$81.9\pm0.3$&$93.6\pm0.2$&$87.6\pm0.5$&$89.6\pm0.3$&$88.8\pm0.7$&$97.1\pm0.2$&$98.0\pm0.3$\\
    \bottomrule
   \end{tabular}
\end{table*}	
\begin{table*}[htpb]
   \centering
   \caption{The classification accuracies (\%) of our MCNN and six comparative methods tested on University of Pavia}
   \label{tab2}
   \begin{tabular}{p{0.15\linewidth}<{\centering}p{0.08\linewidth}<{\centering}p{0.08\linewidth}<{\centering}p{0.08\linewidth}<{\centering}p{0.08\linewidth}<{\centering}p{0.08\linewidth}<{\centering}p{0.08\linewidth}<{\centering}p{0.08\linewidth}<{\centering}}
   \toprule
Class&SVM&EPF&3DCNN$_1$&3DCNN$_2$&DHCNet&SSRN&MCNN\\  
    \midrule
   Asphalt&$98.0\pm1.3$&$99.0\pm0.1$&$92.2\pm1.2$&$98.2\pm1.0$&$93.4\pm0.6$&$99.1\pm0.4$&$99.6\pm0.3$\\
  Meadows&$98.7\pm0.2$&$99.6\pm0.1$&$98.5\pm0.1$&$97.2\pm1.2$&$98.3\pm1.5$&$99.5\pm0.6$&$99.4\pm0.3$\\
   Gravel&$80.1\pm2.3$&$97.1\pm1.2$&$98.7\pm1.3$&$96.2\pm0.8$&$95.5\pm1.2$&$99.2\pm0.2$&$99.4\pm0.2$\\
   Trees&$86.8\pm0.2$&$99.6\pm0.2$&$99.7\pm0.2$&$91.5\pm0.8$&$98.5\pm0.4$&$99.5\pm0.4$&$99.8\pm0.1$\\
   Metal sheets&$97.5\pm0.4$&$99.3\pm0.2$&$99.8\pm0.1$&$99.6\pm0.3$&$96.5\pm1.1$&$100\pm0$&$99.9\pm0.1$\\
   Bare soil&$85.2\pm3.5$&$94.7\pm1.8$&$99.8\pm0.2$&$97.8\pm0.4$&$97.5\pm5.2$&$98.8\pm0.8$&$99.8\pm0.2$\\
  Bitumen&$72.2\pm4.3$&$100\pm0$&$84.5\pm0.8$&$96.4\pm0.9$&$98.6\pm0.6$&$99.3\pm0.5$&$98.8\pm0.5$\\
  Bricks&$87.5\pm1.2$&$93.8\pm0.8$&$99.2\pm0.5$&$87.8\pm0.5$&$97.5\pm1.2$&$98.5\pm0.4$&$98.9\pm0.2$\\
 shadows&$99.7\pm0.3$&$100\pm0$&$99.4\pm0.3$&$100\pm0$&$98.2\pm0.3$&$100\pm0$&$98.4\pm0.4$\\
  OA&$93.5\pm0.4$&$98.3\pm0.3$&$97.5\pm0.4$&$96.3\pm0.6$&$97.2\pm0.8$&$99.3\pm0.2$&$99.5\pm0.2$\\
  AA&$89.5\pm0.3$&$98.1\pm0.2$&$96.9\pm0.5$&$96.0\pm0.6$&$97.1\pm0.5$&$99.3\pm0.2$&$99.3\pm0.1$\\
  $\kappa\times100$&$91.4\pm0.3$&$97.8\pm0.3$&$96.7\pm0.3$&$95.1\pm0.5$&$96.3\pm0.6$&$99.0\pm0.1$&$99.3\pm0.2$\\
    \bottomrule
   \end{tabular}
\end{table*}	
\begin{table*}[htpb]
   \centering
   \caption{The classification accuracies (\%) of our MCNN and six comparative methods tested on Salinas}
   \label{tab2}
   \begin{tabular}{p{0.15\linewidth}<{\centering}p{0.08\linewidth}<{\centering}p{0.08\linewidth}<{\centering}p{0.08\linewidth}<{\centering}p{0.08\linewidth}<{\centering}p{0.08\linewidth}<{\centering}p{0.08\linewidth}<{\centering}p{0.08\linewidth}<{\centering}}
   \toprule
Class&SVM&EPF&3DCNN$_1$&3DCNN$_2$&DHCNet&SSRN&MCNN\\  
    \midrule
Weeds\_1&$97.6\pm1.0$&$100\pm0$&$80.2\pm4.2$&$100\pm0$&$95.8\pm0.2$&$99.7\pm0.1$&$100\pm0$\\
Weeds\_2&$99.2\pm0.5$&$100\pm0$&$88.2\pm2.5$&$84.1\pm2.1$&$99.6\pm0.2$&$99.3\pm0.1$&$99.4\pm0.1$\\
   Fallow&$91.2\pm1.1$&$95.8\pm0.2$&$99.2\pm0.1$&$92.8\pm0.5$&$99.8\pm0.2$&$99.4\pm0.2$&$99.5\pm0.1$\\
   Fallow\_P&$99.1\pm0.2$&$99.5\pm0.1$&$99.3\pm0.2$&$97.4\pm0.2$&$96.0\pm0.5$&$98.6\pm0.2$&$99.4\pm0.5$\\
   Fallow\_S&$95.8\pm1.1$&$99.4\pm0.3$&$99.5\pm0.5$&$98.3\pm0.8$&$98.7\pm1.0$&$100\pm0$&$98.2\pm0.2$\\
   Stubble&$100\pm0$&$100\pm0$&$99.5\pm0.2$&$98.7\pm0.5$&$99.3\pm0.7$&$98.6\pm0.7$&$99.7\pm0.1$\\
  Celery&$99.1\pm0.1$&$95.4\pm4.4$&$97.3\pm0.7$&$97.2\pm0.2$&$99.6\pm0.2$&$99.0\pm0.5$&$98.8\pm0.3$\\
  Grapes&$78.5\pm4.5$&$90.8\pm0.4$&$90.2\pm0.3$&$91.2\pm0.7$&$96.3\pm1.2$&$99.6\pm0.1$&$99.4\pm0.2$\\
  Soil&$99.2\pm0.2$&$99.3\pm0.3$&$100\pm0$&$100\pm0$&$99.6\pm0.3$&$92.0\pm0.4$&$99.7\pm0.1$\\
  Corn&$86.8\pm1.2$&$92.2\pm0.4$&$96.2\pm0.4$&$99.2\pm0.1$&$88.8\pm0.2$&$100\pm0$&$100\pm0$\\
  Lettuce\_4wk&$86.5\pm1.2$&$97.2\pm0.3$&$98.2\pm1.5$&$99.6\pm0.1$&$100\pm0$&$100\pm0$&$100\pm0$\\
  Lettuce\_5wk&$98.2\pm2.3$&$96.5\pm3.4$&$92.3\pm0.4$&$83.8\pm1.2$&$86.2\pm4.2$&$92.1\pm1.3$&$97.2\pm1.2$\\
 Lettuce\_6wk&$93.2\pm1.3$&$98.4\pm1.2$&$98.5\pm1.6$&$82.9\pm4.4$&$98.1\pm1.4$&$99.2\pm0.2$&$99.7\pm0.2$\\
  Lettuce\_7wk&$87.5\pm0.4$&$99.1\pm0.1$&$78.4\pm3.4$&$87.2\pm3.3$&$98.2\pm1.1$&$99.2\pm0.2$&$99.7\pm0.2$\\
  Vinyard\_U&$75.8\pm5.6$&$85.1\pm2.4$&$98.1\pm1.6$&$93.5\pm0.2$&$83.3\pm1.5$&$99.2\pm0.2$&$98.8\pm0.1$\\
  Vinyard\_T&$98.0\pm0.3$&$95.4\pm1.2$&$99.1\pm0.5$&$99.5\pm0.1$&$99.6\pm0.2$&$98.2\pm0.2$&$99.8\pm0.1$\\
  OA&$89.9\pm0.6$&$94.7\pm0.4$&$94.9\pm0.5$&$94.3\pm0.4$&$95.3\pm0.3$&$98.3\pm0.2$&$99.3\pm0.2$\\
  AA&$92.8\pm0.5$&$96.5\pm0.3$&$94.6\pm0.6$&$94.1\pm0.3$&$96.2\pm0.4$&$98.4\pm0.2$&$99.3\pm0.1$\\
   $\kappa\times100$&$88.9\pm0.8$&$94.1\pm0.4$&$94.3\pm0.4$&$93.7\pm0.5$&$94.7\pm0.3$&$98.0\pm0.3$&$99.2\pm0.2$\\
    \bottomrule
   \end{tabular}
\end{table*}	

We have compared the computational time of four neural network methods with our method in Table VIII-X. In these methods, DHCNet has employed PCA for data preprocessing, SSRN does not employ a data preprocessing method for the special architecture. The original method of SSRN aims at extracting spectral-spatial feature simultaneously. They use the HSI patch of size $13\times13\times200$ as input.
The 3DCNN$_1$ and 3DCNN$_2$ also skip the data preprocessing step. Because of the same reason in SSRN, these two methods also emphasize the spectral-spatial feature extraction. To make the test fair, we can apply PCA on these methods in the same manner to reduce the spectral bands. If we employ PCA to reduce the feature dimension from $13\times13\times200$ to $7\times7\times40$, the spectral-spatial structure may be destroied because a patch is flattened as a vector. Thus, we only use PCA to reduce the spectral bands to 40 for 3DCNN$_1$, 3DCNN$_2$ and SSRN.

Deep learning methods are time consuming. There are three points may affect the running time: convolution operation, the number of convolutional layers and the number of training epochs. We make improvements upon these three points. The input of our convolutional layer is smaller than that of the original HSI patch, which results in less convolution operations. Meanwhile, we design the mapping layer aiming at reducing the spectral and spatial redundancy in HSI. Therefore, less convolutional layers are needed in our MCNN. Our effective architecture can also save the number of training epochs. Since the training loss in our network decreases fast, merely 30 epochs or less are sufficient. In summary, the time of convolutional operations in one convolutional layer is saved, the number of convolution layers is reduced and the training process is more effective in our MCNN architecture. 
\begin{table}[htpb]
   \centering

   \caption{The training time (s) on Indian Pines by deep learning methods: SSRN, DHCNet, 3DCNN$_1$, 3DCNN$_2$ and MCNN}
   \label{tab2}
   \begin{tabular}{p{0.25\columnwidth}<{\centering}p{0.08\columnwidth}<{\centering}p{0.08\columnwidth}<{\centering}p{0.08\columnwidth}<{\centering}p{0.08\columnwidth}<{\centering}p{0.08\columnwidth}<{\centering}}
   \toprule
Indian Pines&3DCNN$_1$&3DCNN$_2$&SSRN&DHCNet&MCNN\\
  \midrule
Training time for each epoch (s)&2.08&2.61&$\bm{0.74}$&-&$0.77$\\	Epochs&30&30&200&-&30\\	
Preprocessing time (s)&55.80&56.12&55.52&30.04&$\bm{0.26}$\\	
Training time in total (s)&118.24&134.52&203.47&393.59&$\bm{23.15}$\\  	
    \bottomrule
   \end{tabular}
\end{table}	

\begin{table}[htpb]
   \centering
   \caption{The training time (s) on University of Pavia by deep learning methods: SSRN, DHCNet, 3DCNN$_1$, 3DCNN$_2$ and MCNN}
   \label{tab2}
   \begin{tabular}{p{0.25\columnwidth}<{\centering}p{0.08\columnwidth}<{\centering}p{0.08\columnwidth}<{\centering}p{0.08\columnwidth}<{\centering}p{0.08\columnwidth}<{\centering}p{0.08\columnwidth}<{\centering}}
   \toprule
University of Pavia&3DCNN$_1$&3DCNN$_2$&SSRN&DHCNet&MCNN\\
  \midrule
Training time for each epoch (s)&5.81&8.03&$\bm{1.89}$&-&3.34\\		
Epochs&30&30&200&-&30\\		
Preprocessing time (s)&72.34&73.46&73.16&56.25&$\bm{0.14}$\\	
Training time in total (s)&246.87&314.26&452.63&389.15&$\bm{100.21}$\\  	
    \bottomrule
   \end{tabular}
\end{table}	
\begin{table}[htpb]
   \centering
   \caption{The training time (s) on Salinas by deep learning methods: SSRN, DHCNet, 3DCNN$_1$, 3DCNN$_2$ and MCNN}
   \label{tab2}
   \begin{tabular}{p{0.25\columnwidth}<{\centering}p{0.08\columnwidth}<{\centering}p{0.08\columnwidth}<{\centering}p{0.08\columnwidth}<{\centering}p{0.08\columnwidth}<{\centering}p{0.08\columnwidth}<{\centering}}
   \toprule
Salinas&3DCNN$_1$&3DCNN$_2$&SSRN&DHCNet&MCNN\\
  \midrule
Training time for each epoch (s)&3.35&5.35&1.25&-&$\bm{1.04}$\\		
Epochs&30&30&200&-&30\\	
Preprocessing time (s)&64.25&65.74&63.51&48.41&$\bm{0.22}$\\		
Training time in total (s)&164.86&226.47&314.28&416.24&$\bm{31.53}$\\  	
    \bottomrule
   \end{tabular}
\end{table}	

\subsection{The Efficiency of Our Mapping Layers}
To further verify the advantage that our mapping layer can extract better spectral-spatial feature, we test our MCNN against three specified situations. 
\begin{itemize}

\item[1] Mapping layers in our architecture are replaced by PCA transform. The PCA is employed as a preprocessing method on training dataset and testing dataset separately. The spectral bands are reduced to 40. 

\item[2] Mapping layers are removed and 3D-convolutional layers are directly applied to classify HSI raw data. The input of MCNN is the original HSI patches.

\item[3] Mapping layers are replaced by Tucker decomposition (TD). We apply TD on each training patche and testing patche. The sizes of core tensor are the same with the output of our last mapping layer. 
\end{itemize}

We exam test three situations on Indian Pines. The overall accuracy (OA) average accuracy (AA) and kappa coefficient, preprocessing time and total training time results are listed in Table XI.

\begin{table}[htpb]
 \centering

   \caption{The classification performance of three situations compared with our mapping layers}

   \label{tab2}
   \begin{tabular}{p{0.2\columnwidth}<{\centering}p{0.1\columnwidth}<{\centering}p{0.1\columnwidth}<{\centering}p{0.1\columnwidth}<{\centering}p{0.15\columnwidth}<{\centering}}
   \toprule
Indian Pines&PCA&Raw data&TD&Mapping layers\\
  \midrule
OA(\%)&95.65&88.86&96.34&98.53\\
AA(\%)&94.24&87.46&95.87&97.09\\
$\kappa$&0.9532&0.8785&0.9612&0.9769\\
Preprocessing time (s)&55.80&-&68.34&0.26\\
Total time (s)&118.24&266.25&90.12&23.15\\		 	
    \bottomrule
   \end{tabular}
\end{table}	

It can be seen that our proposed method performs best. In Situation 1, we replace our mapping layers with PCA, its classification accuracy results are nearly 3\% lower than our method. The classification accuracy results show that PCA cannot extract spectral-spatial features as well as our mapping layers. In Situation 2, the mapping layers are removed to test the feature extraction performance. The classification accuracy results of Situation 2 are the lowest among four compared methods. It is easy to understand that mapping layers can extract good spectral-spatial features for classification, which is obviously better than original raw data. In Situation 3, mapping layers are replaced by TD. TD can well decompose the input patch, however, each input patche is decomposed separately to obtain core tensors. Thus, the common spectral-spatial features are not well extracted and the classification accuracy is not competitive with ours. Besides, the training time is also larger than our method.

The preprocessing time of our mapping layers is negligible, because our mapping layers are obtained from the averaged training patches. Since there are $K$ patches of sizes $M\times N\times Z$, PCA is applied on a matrix of size $KMN\times Z$, which needs  $Z^2KMN$ multiplications. TD is applied on $K$ tensors of size $M\times N\times Z$, which needs $(MZ+MN+ZN)MNZK$ multiplications. While our method is applied on a tensor of sizes $M\times N\times Z$, which only needs $(MN)^2Z+(ZN)^2M+(MZ)^2N$ multiplications. Usually, $K$ is a large number in HSI processing, which is saved in our method. Thus, the training time of our mapping layers (which is also the preprocessing time in Table XI) is negligible.

In summary, our mapping layers can extract better spectral-spatial feature which helps to improve the classification accuracy performance, mapping layers can also save the training time of our MCNN architecture.

The disadvantage of our MCNN is that it only avoids the gradient vanishing problem, this problem can only be sovled by residual network so far. Meanwhile, our MCNN cannot extract features of different scales. These two problems will be investigated in our future work.

\section{Conclusion}
In this paper, we have presented a neural network architecture of CNN with mapping layers (MCNN) for spectral-spatial feature extraction and HSI classification. The proposed MCNN architecture, which contains mapping section designed for spectral and spatial redundancy reduction, convolutional section to extract abstract features and fully connected section to classify the features, has alleviated the spectral-spatial feature extraction and avoided the decreasing accuracy phenomenon. It is worth noting that our proposed mapping layer uses the multilinear algebra to map the input HSI patch into lower dimensional subspace and maintains the spectral and spatial structure. Meanwhile, most energy of the patch is preserved. 
Moreover, our MCNN architecture simplifies the spectral-spatial feature extraction and improves the quality of spectral-spatial feature. It not only avoids the problem that classification accuracy decreases with the increasing number of convolutional layers, but also saves the training time of network. Furthermore, we employ the 3-D convolutional kernel to extract the spectral-spatial feature in convolutional layers.
Experiment results demonstrated that our proposed MCNN performs consistently with the highest overall accuracy for all three commonly used HSI datasets with various challenges.

\ifCLASSOPTIONcaptionsoff

\fi

\bibliographystyle{IEEEtran}
\bibliography{IEEEabrv}

%



\begin{IEEEbiographynophoto}{Rui Li}
received the B. S. and M. S. degree from Xidian University, Xi'an, P. R. China, in 2012 and 2015 respectively. 
He is currently working towards the Ph. D. degree in School of Electronic and Information Engineering at Xi'an Jiaotong University, Xi'an, P. R. China. His research interests include vector quantization and hyper-spectral image processing.
\end{IEEEbiographynophoto}
\begin{IEEEbiographynophoto}{Zhibin Pan}
received the B. S. degree in Information and Telecommunication Engineering in 1985 and the M. S. degree in Automation Science and Technology in 1988 from Xi'an Jiaotong University, P. R. China, respectively. He received the Ph.  D. degree in Electrical Engineering in 2000 from Tohoku University, Japan. He is a professor in the Department of Information and Telecommunication Engineering, Xi'an Jiaotong University, P. R. China. His current research interests include image compression, multimedia security and object recognition.
\end{IEEEbiographynophoto}
\begin{IEEEbiographynophoto}{Yang Wang}
received his B. S. degree from Xi'an Jiaotong University, Xi'an, P. R. China, in 2010.
He is currently working towards the Ph. D. degree in School of Electronic and Information Engineering at Xi'an Jiaotong University, Xi'an, P. R. China. His research interests include image coding and image processing.
\end{IEEEbiographynophoto}
\begin{IEEEbiographynophoto}{Ping Wang}
received her B. S., M. S. and Ph. D. degrees from Xi'an Jiaotong University, Xi'an, P. R. China, in 1999, 2002, and 2011, respectively, all in information and communication engineering. She is currently an associate professor with the School of Electronic and Information Engineering, Xi'an Jiaotong University. Her research interests include image processing, video coding and video analysis.
\end{IEEEbiographynophoto}





\end{document}